\documentclass[aps,prc,groupedaddress,preprint]{revtex4}
\def\ohalf{{\textstyle{1\over 2}}}
\def\half{{\textstyle{1\over 2}}}

\def\osix{{\textstyle{1\over 6}}}

\def\thalf{{\textstyle{3\over 2}}}
\def\fourth{{\textstyle{1\over 4}}}

\def\M{{\cal M}}
\def\P{{\sf P}}
\def\R{{\cal  R}}
\def\D{{\cal D}}
\def\I{{\cal I}}

\def\MD{{M^*}}

\newcommand{\beq}{\begin{equation}}
\newcommand{\be}{\begin{equation}}
\newcommand{\eeq}{\end{equation}}
\newcommand{\ee}{\end{equation}}
\newcommand{\beqa}{\begin{eqnarray}}
\newcommand{\bea}{\begin{eqnarray}}
\newcommand{\eeqa}{\end{eqnarray}}
\newcommand{\eea}{\end{eqnarray}}
\newcommand{\bra}[1]{\langle {#1} |}                        % < |
\newcommand{\ket}[1]{| {#1} \rangle}

\usepackage{epsfig}

\begin{document}

\title{Baryon Form Factors of Relativistic Constituent-Quark Models}

\author{B. Juli\'a-D\'{\i}az}
\email[]{Bruno.Julia@helsinki.fi}
\author{D. O. Riska}
\email[]{riska@pcu.helsinki.fi}
\affiliation{Helsinki Institute of Physics and Department of Physical Sciences, POB 64,
00014 University of Helsinki, Finland}

\author{F. Coester}
\email[]{coester@theory.phy.anl.gov}
\affiliation{Physics Division, Argonne National Laboratory,
Argonne, IL 60439, USA}

\date{\today}

\begin{abstract}

The electromagnetic and axial form factors of the nucleon and its lowest 
positive parity excitations, the $\Delta(1232)$ 
and the $N(1440)$, are calculated with constituent-quark models that
are specified
by simple algebraic representations of the mass-operator eigenstates.
Poincar\'e covariant current operators are generated by the 
dynamics from single-quark currents that are covariant under a 
kinematic subgroup. The dependence of the calculated form factors on the 
choice of kinematics and on the gross features of the wave functions is
illustrated for instant-form, point-form, and front-form kinematics.
A simple algebraic form of the orbital
ground state wave function, which depends on two parameters, 
allows a fair description of all the form factors over 
the empirically accessible range,
although with widely different choices of the parameters, which determine the
range and shape of the orbital wave function.
The neutron electric form factor requires additional features,
for instance the presence of mixed symmetry $S-$state component
with 1 -- 2 \% probability in the ground state wave function.
Instant and front form kinematics demand a spatially extended
wave function, whereas in point form kinematics the form
factors may be described with a quite compact wave function. 
\end{abstract}

\maketitle

\section{Introduction}

\label{intro}

The baryon states of constituent-quark models are represented by 
functions of three quark coordinates, spin, flavor and color 
variables. The Hilbert space of states is the tensor product of 
three irreducible Poincar\'e representations.
The bound state wave functions represent vectors in the 
representation space of the little group of the Poincar\'e transformations. 
A calculation of baryon form factors requires consistent representations
of the current-density operators and the baryon states. 
While current density operators are represented by functions of 
quark velocities and spinor indices, the baryon states are represented by 
eigenfunctions of the mass operator, which are functions of 
internal momenta, $\vec k_i$, and spin variables. The relation between 
the two representations depends on the ``form of kinematics'', which specifies a 
kinematic subgroup of the Poincar\'e group.

Following Bakamjian and Thomas \cite{bakam} the baryon states may be 
represented by eigenfunctions of the mass operator, 
$\M$, the spin operator, $j^2, j_z$,
and three kinematic 
operators which, together with the mass operator, specify the 
four-momentum. The mass operator commutes with these operators and is 
independent of their eigenvalues, which therefore may be treated as parameters.
Relevant examples are the velocity $\vec v$, the three-momentum $\vec P$,
and the light-front momentum ${\bf P}:=\{P^+, P_\perp\}$, 
with the four-momentum represented respectively by
\be
P = \M\left\{\sqrt{1+|\vec v|^2},\vec v\right\}\;,\qquad
P=\left\{\sqrt{|\vec P|^2+\M^2},\vec P\right\}\;,\qquad
P=\left\{{\M^2+P_\perp^2\over P^+},{\bf P}\right\}\;.
\label{eq1}
\ee
The choice of these kinematic parameters implies the ``form of kinematics'',
which is the choice of a kinematic subgroup of the Poincar\'e group~\cite{coe02}.
The Poincar\'e representations of the kinematic subgroup are independent 
of the mass operator. In particular, they are the same as the representations 
of free quarks with mass operator $\M_0$. The kinematic subgroups, which 
correspond to the momentum representations in Eq.~(\ref{eq1}) are the Lorentz 
group, $SO(1,3)$, the Euclidean group in three dimensions $E(3)$ (translations 
and rotations at a fixed time), and the symmetry group of the null-plane $n\cdot x=0,\; n^2=0$. 
Following Dirac's seminal paper~\cite{dirac}, the three forms of kinematics are 
referred to as point form, instant form and front form.

The relations between the internal momenta and spin variables to the quark 
velocities and spinor variables depend on boost parameters which are
$\vec v$, $\vec P$ and $P_\perp/ P^+$, 
with Lorentz kinematics, instant kinematics and light-front kinematics,
respectively. Poincar\'e covariant current-density operators can be generated 
by the dynamics from current operators that are covariant under the kinematic 
subgroup only. Employment of free-quark currents for that purpose leads 
to different current operators in the different forms of kinematics. The 
quark masses enter as essential scale parameters of these current operators. 

While the mass operator of conventional quark models, e.g.~\cite{boffi}, also
depends on the quark masses, baryon spectra of confined quark may be represented by 
mass operators that are independent of quark masses~\cite{coe98}. 
Eigenfunctions of such mass operators can be consistent with empirical nucleon 
form factors~\cite{coe03}. Nucleon models constructed in this manner 
depend only on one scale that can be varied by unitary transformations.
Form factors are dimensionless functions of the invariant velocity difference
$\eta:=1/4(v_{out}-v_{in})^2$, and the mass ratio $M_{out}/M_{in}$. With 
Lorentz kinematics $\eta$ is a kinematic quantity and the form factors are 
relatively insensitive to unitary scale transformations of the wave function. 
Relations to momentum transfers involve the baryon masses. With instant and 
light-front kinematics momentum transfers are kinematic, since for
these there is kinematic
translation covariance in 3 or 2 space dimensions respectively.

Poincar\'e covariant state vectors of few-body systems are represented by 
equivalence classes of functions~\cite{polyzou} and there is no relation 
of a particular representation to wave functions defined by matrix elements 
of field operators~\cite{desplanques, brodsky}.

The purpose of this paper is to explore the dependence of the baryon elastic 
and transition form factors on the representation of the baryon mass operator 
and on the form of kinematics used in the construction of the current operators. 
For that purpose we assume a Bakamjian-Thomas representation of the baryon 
states and generate current density operators 
from simple quark currents that 
are covariant under the kinematic subgroup only. Non-Bakamjian-Thomas 
representations of the baryon states are equivalent by unitary transformations, 
which modify the representation of the quark currents~\cite{polyzou}.

The mass operator is constructed in a simple spectral representation,
which is 
independent of quark masses:
\bea
&&\bra {\sigma_1,\sigma_2,\sigma_3,\vec k_1,\vec k_2,\vec k_3}\M
\ket{\sigma_3',\sigma_2',\sigma_1',\vec k'_3, \vec k'_2,\vec k'_1}\cr
&&=\sum_{n,j,\sigma} \phi_{n,j,\sigma}
(\sigma_1,\sigma_2,\sigma_3,\vec k_1,\vec k_2,\vec k_3)M_{n,j}
\phi_{n,j,\sigma}
(\sigma_1',\sigma_2',\sigma_3',\vec k_1',\vec k_2',\vec k_3')^*\, ,
\label{MASS}
\eea
with the restriction to the nucleon, the $\Delta(1232)$ and the $N(1440)$.
Generalization to other states is straightforward. A two-parameter 
family of algebraic functions is employed, which allow variations of the range and the 
shape of the function. Using hyperspherical coordinates, the spatial wave 
function of the $N(1440)$ baryon is constructed with a single node  to be orthogonal to 
the ground state. For a satisfactory description of 
the electric form factor of the neutron a small admixture of $\sim$
1--2 \% of a mixed symmetry $S-$state is included in the ground state wave function.

For the quark currents the same structureless spinor currents are employed
throughout. Variations of this input are beyond the scope of this article.
The quark velocities are related to the internal momenta by boost relations,
which depend on the choice of the kinematics with significant qualitative 
and quantitative consequences. With point-form and light-front 
kinematics different 
quark velocities are related by kinematic Lorentz transformations.
With instant-form kinematics there is no kinematic relation 
between different quark velocities.
With light-front and instant form kinematics translation covariance emphasizes 
the spatial extent, $r_0$, of the wave function. 

When the spatial extent of the wave function is scaled unitarily 
to zero, the calculated form factors become independent of 
momentum transfer in both instant and front form kinematics. In contrast 
point form kinematics has a non-trivial limit, when the spatial extent of 
the wave function is scaled to zero. In this ``point limit'' the calculated 
form factors depend on the functional form of the wave function, and when
$\eta\gg 1$ decrease with an inverse power of the momentum transfer.
The falloff power is determined by the current operator and is independent 
of the wave function \cite{coe03}. 

The present paper is organized in the following way. In Section II 
the model independent relations of covariant current matrices to invariant 
form factors are summarized.
Section III contains the description of the baryon
model specified 
by a mass operator and kinematic quark currents. Section IV contains a detailed
description of the integrals which need to be evaluated after summation over 
spin and flavor indices. Numerical results are presented in Section V. 
A concluding discussion is given in Section VI.

\section{Current-density operators and form factors}

The definition of form factors depends on the Poincar\'e covariance of 
the current density operators $I^\mu(x)$ and the basis states
 $\ket{M,v,j ,j_z}$.
The current density
\be 
I^\mu(x)= e^{i P\cdot x} I^\mu(0) e^{-i P\cdot x}\; ,
\ee
satisfies the Lorentz covariance relations
\be
U^\dagger(\Lambda) I^\mu(x) U(\Lambda) = \Lambda^\mu\,_\nu
I^\nu(\Lambda^{-1} x)\; .
\ee
By definition the spin operator $\vec j$ is related to the Lorentz 
generators $J^{\mu\nu}$ by 
\be
\{0,\vec  j\} := B^{-1}(v)\,w \qquad\mbox{with}\qquad
 w_\tau:= \half v^\nu J^{\rho\sigma}\epsilon_{\nu\rho\sigma\tau}
\label{SPIN}
\ee
where the boost operator $B(v)$ is an operator valued Lorentz transformation 
with the defining property: 
\be
B(v)\{1,0,0,0\}=v\; .
\label{VEL}
\ee 
It follows from the definition (\ref{SPIN}) and the Lorentz covariance 
of the velocity operator,
$ U^\dagger(\Lambda)\, v\, U(\Lambda)= \Lambda v $, that
the spin operator transforms according to
\be
U^\dagger(\Lambda)\, \vec j\, U(\Lambda)=
{\cal R}_W(\Lambda,v) \vec j\;  ,
\ee
with the Wigner rotations ${\cal R}_W(\Lambda,v)$ defined by
\be
{\cal R}_W(\Lambda,v):=B^{-1}(\Lambda v)\Lambda B(v) \,.
\label{RWIG}
\ee
The basis states transform according to
\be
U(\Lambda)\ket{M,v,j,\sigma} = \sum _{\sigma'}
\ket{M,\Lambda v,j,\sigma'} D^j_{\sigma',\sigma}[R_W(\Lambda, v)]\;.
\ee

With definite initial and final velocities and masses, $v_a, M_a$ and 
$v_f, M_f$ the form factors are determined by invariant reduced matrix elements 
of the currents.
They are dimensionless functions of $\eta$:
\be 
\eta:= \fourth(v_f-v_a)^2\; , \qquad  -\fourth(v_f+v_a)^2= 1+\eta\; ,
\ee
and the baryon masses. The relation between the invariant momentum transfer
and $\eta$ is:
\be
Q^2:= (M_f v_f-M_av_a)^2= 4M_f M_a \eta -(M_f-M_a)^2\; .
\ee
In practice the dynamics generates the current operators from
kinematic currents that are covariant under a subgroup only.
It is therefore important to define basis states such that the Wigner
rotations of kinematic transformations are kinematic. For Lorentz and 
instant kinematics canonical boosts satisfy this requirement. For any 
rotation $\R$ the corresponding Wigner rotation satisfies $\R_W(\R,v)=\R$ 
and for rotationless Lorentz transformations in the direction
of $\vec v$ the Wigner rotations reduce to the identity. Light-front kinematic 
requires null-plane boosts defined such that the Wigner rotations 
of null-plane boosts are the identity.

\subsection{Electromagnetic  form factors}

Lorentz and instant kinematics share the subgroup of rotations 
about the direction of the velocities, which suggests the use 
of canonical spins and the separation of the conserved current density, 
$I^\mu(0)$, into ``electric'' 
and ``magnetic'' currents which are projections of the 
current into the plane defined by the velocities $v_a$ and $v_f$
and the projection perpendicular to that plane. 

Current conservation, $(M_f v_f-M_a v_a)\cdot I(0)=0$, implies that the
electric current be a linear combination of the velocities multiplied 
by a single invariant operator $\I_e$. It is satisfied by the 
expression.
\be
I^\mu_e(0)= \left\{ {M_f+M_a\over \sqrt {4 M_f M_a}}
{v^\mu_f+v^\mu_a\over 2\sqrt{1+\eta}} +
{M_f-M_a\over\sqrt{4M_fM_a}}
{v^\mu_f-v^\mu_a\over 2\sqrt{\eta} }\sqrt{{1+\eta\over \eta}}\right\}\I_e \, ,
\label{ELEC}
\ee
which implies
\be
\I_e:=-{(M_a v_f+M_f v_a)\over 2\sqrt{1+\eta}\sqrt{M_a M_f}}
\cdot I(0)\, .
\ee

The choice of coordinate axes is a matter of convenience. When the z-axis is 
in the direction of the velocities, the components of the magnetic current are
\be
I_m(0)=\{0,\I_{mx}(\eta),\I_{my}(\eta),0\}\, .
\label{MAG}
\ee
Electric and magnetic form factors are invariant matrix elements of the 
expressions (\ref{ELEC}) and (\ref{MAG}).
Both $\I_e$ and $\vec \I_m$ are invariant under rotationless Lorentz 
transformations in the z-direction and, with canonical boosts, 
the corresponding Wigner rotations are the identity.
Form factors can thus be defined by invariant canonical-spin matrix elements.

The elastic form factors of the nucleon are defined by:
\bea
G_E(\eta)&:=&\bra{\half} \I_e(\eta)\ket{\half} _c= \bra{-\half} 
\I_e(\eta)\ket{-\half}_c\; ,\\ \nonumber
G_M(\eta)&:=&{1\over \sqrt{\eta}}  
\bra{\half}\I_{mx}(\eta)\ket{-\half}_c=-{1\over \sqrt{\eta}}  
\bra{-\half}\I_{mx}(\eta)\ket{\half}_c  \;. \label{GEGM}
\eea
The magnetic form factor for the transition between a spin
1/2 and a spin 3/2 state can be defined by:
\beq  
G_{M\thalf}(\eta):={1\over \sqrt{\eta}}  
\bigg[  \bra{\half,\half}\I_{mx}(\eta)\ket{\thalf,-\half}_c 
      +\sqrt{3} \bra{\half,-\half}\I_{mx}(\eta)\ket{\thalf,-\thalf}_c \bigg] 
\, .
\eeq
With Lorentz kinematics all Lorentz transformations are kinematic and
the choice of a ``frame'', for the component of the velocities, for instance 
\be
v_f=\left\{\sqrt{1+\eta},0,0,\sqrt {\eta}\right\}\, ,
\qquad v_a=\left\{\sqrt{1+\eta},0,0,
-\sqrt{\eta}\right \}\;,
\label{BREIT}
\ee
is a matter of convenience.

The light-front kinematic subgroup leaves the null-plane,
$n\cdot x=0,\; n^2=0$ invariant. For space-like momentum transfer, $Q^2>0$,
the null vector $n$ can be chosen such that $P_f^+=P_a^+$.
Thus the operator $I^+(0)/P^+$ and the null-plane spin are invariant 
under the kinematic subgroup. Form factors can be defined by the
dimensionless current matrix
\be
\I^+ := {\sqrt{M_fM_a}\over P^+} I^+(0) \, .
\ee
For a given velocity $v$ the light-front spin is related to the canonical 
spin by the Melosh rotation $\R_M(v):= B_c(v)B^{-1}(v)$. With the null 
vector $n=\{-1,0,0,1\}$ the condition $Q^+=0$ requires velocities at an angle 
$\alpha$ relative to the $z-$axis that depend on 
$\eta$ and the ratio $M_f/M_a$,
\be
\cos \alpha := {M_a-M_f\over M_f+M_a}\sqrt{{1+\eta\over \eta}}\; .
\ee

With instant kinematics the kinematic subgroup, which leaves some 
time-like vector $n$ invariant, does not include rotationless Lorentz 
transformations. The kinematic boost parameters $\vec  P_f$ and $ \vec P_a$
are related kinematically only when the vector $n$ is chosen in the 
direction of $M_a v_a +M_f v_f$. Then, $\vec P_f= - \vec P_a=\half \vec Q$. 
A consistent calculation of the form factors with instant-form kinematics
requires the same ``frame'', $\vec P_f= - \vec P_a$, for both elastic and 
transition form factors. It follows that the momentum transfer $\vec Q^2$
is a function of $Q^2$ and the baryon masses,
\beq 
\vec Q^2 =Q^2- {[(P_f+P_a)\cdot  Q]^2\over (P_f+P_a)^2}=
Q^2 + {(M_f^{2}-M_a^2)^2\over Q^2 + 2(M_f^{2}+M_a^2)}\, .
\label{Qs}
\eeq
The velocities are then
\bea
v_f &=& \{ \sqrt{1+\eta} \cosh \chi +\sqrt{\eta}\sinh \chi,0,0, 
\sqrt{\eta}\cosh \chi +\sqrt{1+\eta}\sinh\chi\}\, ,\cr
v_a&=& \{ \sqrt{1+\eta} \cosh \chi -\sqrt{\eta} \sinh \chi,0,0, 
-\sqrt{\eta}\cosh \chi +\sqrt{1+\eta}\sinh\chi\}\, ,
\label{INVEL}
\eea
where
\be
{\sinh \chi\over \cosh \chi}
:= {M_a-M_f\over M_f+M_a}\sqrt{\eta\over 1+\eta}\; .
\ee

The evaluation of the null-spin matrix $\I^+$ and the canonical-spin 
matrices $\I_e,\I_m$ necessarily requires different orientations of the 
velocities. The relations of the canonical spin matrices $\I_e,\I_m$ to the 
null-plane-spin matrix $\I^+$ can be conveniently established by the relations 
of the spin representations to spinor representations provided by the spinor 
representations of light-front boosts $u_f(v)$ and canonical boosts $u_c(v)$,
which for spin $\half$ are:

\bea
u_f(v)&:=& {\alpha_\perp\cdot v_\perp +\beta + v^+\over \sqrt{v^+}}\, {\alpha^+\over2}
{1+\beta\over 2}\, ,\\\cr
u_c(v)&:=& {\vec \alpha\cdot \vec v + 1 +v^0\over \sqrt{ 2(1+v^0)}}{1+\beta\over 2}\; .
\eea
The light-front-spin matrix
\bea
\I^+ &=& u_f(v_f) \left( \gamma^+ F_1 
-\half 
\left[ {Q\cdot \gamma\over \sqrt{4M_aM_f}},\gamma^+\right]F_2\right)u_f(v_a)
{1\over\sqrt{v_f^+v_a^+}}\cr\cr
&=& F_1-\imath \sigma_y \, \sqrt{Q^2\over 4 M_fM_a} F_2\; ,
\eea
with
\be
v_f^+= {P^+\over M_f}\; ,\quad v_{f\perp}= {P_\perp\over M_f}\; ,\qquad
v_a^+= {P^+\over M_a}\; ,\quad v_{a\perp}= {-P_\perp\over M_a}\; ,
\label{VNULL}
\ee
is related to the canonical-spin matrices $\I_e$ and $\vec \I_m$, by
\bea
\I_e&=&\bar u_c(v_f) \left( \left[-v_e\cdot \gamma+(\gamma\cdot Q){v_e\cdot Q\over Q^2}\right] F_1 +
 \half \left[ {Q\cdot \gamma\over \sqrt{4M_aM_f}},v_e\cdot
 \gamma\right]F_2\right)  u_c(v_a)\, ,\cr\cr \cr
\I_{mx}&= & \bar u_c(v_f) \left( \gamma_x F_1
 - \half \left[ {Q\cdot \gamma\over \sqrt{4M_aM_f}},\gamma_x\right]F_2\right)  
u_c(v_a)\, .
\eea
Here 
\beq
v_e:={(M_a v_f + M_f v_a)\over \sqrt{1+\eta}\sqrt{4 M_a M_f}}\, ,
\eeq
and the z-axis is in the direction of the velocities.  It follows that
\be
G_M = F_1 + {M_f+M_a\over 2 \sqrt{M_fM_a}} F_2 \,.
\ee
The corresponding relations for $\half \to \thalf $ transitions are 
derived in Appendix~\ref{appnd}.

\subsection{Axial Form factors }

Axial form factors of the nucleon are defined by the spinor representation,
\be
{\cal A}^\mu = G_A(\eta) \, \gamma^\mu \gamma_5 + G_P \half (v_f-v_a)^\mu \gamma_5\, .
\ee
The form factors are thus related to canonical-spin matrices with the 
velocities (\ref{BREIT}) by
\bea
\bar  u_c(v_f) {\cal A}_x u_c(v_a)& =&\sqrt{1+\eta}\,G_A\, \sigma_x\, , \nonumber \\
\bar  u_c(v_f) {\cal A}_y u_c(v_a)& =&\sqrt{1+\eta}\,G_A\, \sigma_y\, ,
\label{GA}\\
\bar  u_c(v_f) {\cal A}_z u_c(v_a) &=&(G_A-\eta \, G_P)\, \sigma_z \,.
\label{P} 
\eea
The symmetry relation between $x$ and $y$ components is 
kinematic for Lorentz and instant kinematics. 

Null-plane-plane spin matrices with (\ref{VNULL}) and $n=\{-1,0,0,1\}$ are 
related to the form factors by
\be
{\bar u_f(v_f)\, {\cal A}^+ \, u_f(v_a)\over \sqrt{v_f^+v_a^+}}=   G_A\, \sigma_z\; ,
\ee
and
\be
\imath \sigma_z
\bar u_f(v_f)\, {\cal A}_x\,  u_f(v_a) 
+\bar u_f(v_f)\, {\cal A}_y \, u_f(v_a)= \eta G_P\, \sigma_y \,.
\ee

\section{The baryon model}

\subsection{Specification of the mass operator}

For the constituent-quark models under consideration the mass operator $\M$ is
defined by Eq.(\ref{MASS}) with the empirical baryon masses and eigenfunctions 
of $\M$ represented by functions of the form
$\phi_{j,\sigma}(\vec k_1,\vec k_2,\vec k_3;\sigma_1,\sigma_2,\sigma_3)$, 
for which an inner product is defined as
\bea
(\phi_{j',\sigma'},\phi_{j,\sigma})&=&\sum_{\sigma_1,\sigma_2,\sigma_3}
\int d^3 k_1 \int d^3 k_2 \int d^3 k_3 
\delta\left(\vec k_1+\vec k_2+\vec k_3\right)\cr\cr
&\times&
\phi_{j',\sigma'}(\vec k_1,\vec k_2,\vec k_3;\sigma_1,\sigma_2,\sigma_3)^*
\phi_{j,\sigma}(\vec k_1,\vec k_2,\vec k_3;\sigma_1,\sigma_2,\sigma_3)
=\delta_{j',j}\delta_{\sigma',\sigma} \, .
\eea
These functions also depend on flavor and color variables, which are not 
shown explicitly. They are independent of the kinematic parameters 
$\vec v, \; \vec P$ or $P^+,P_\perp$ of the three different forms of kinematics.

In the representation (\ref{MASS}) invariance under rotations is 
necessary and sufficient for the Lorentz invariance of the mass operator.
In this representation the wave function is independent of ``frames''.

The single-baryon wave functions under consideration are products
of functions of the color variable that are anti-symmetric under permutations
with permutation symmetric functions of space, spin and flavor variables. The color functions 
play no role in the form factor calculations and will be suppressed.
For the nucleon ($N$), its first radial excitation ($N(1440)$) and 
the first spin-flip resonance ($\Delta(1232)$) simple representations
without spin-orbit coupling are products permutation-symmetric 
spin-flavor functions,
$\chi_{j,\sigma}(\sigma_1,\tau_1,\sigma_2,\tau_2\sigma_3,\tau_3)$,
with invariant functions of the constituent momenta 
$\varphi_i(\kappa^2+q^2)$, where the Jacobi momenta are defined as:
\be
\vec{\kappa}:=\sqrt{2\over 3}\left(\vec{k}_1 - 
{\vec{k}_2+\vec{k}_3\over2}\right) = \sqrt{3\over 2}\vec k_1
= - \sqrt{3\over 2}(\vec k_2+\vec k_3) \;, \qquad
\vec{q}:=\sqrt{1\over2}(\vec{k}_2-\vec{k}_3) \, .
\ee
Under Lorentz transformations the momenta $\vec k_i$ undergo Wigner rotations 
(\ref{RWIG}),
\be
U^\dagger(\Lambda) \vec k_i U(\Lambda)= \R_W(\Lambda,v) \vec k_i\; .
\label{KWIG}
\ee
It follows that the quadratic sum,
\be
\kappa^2+q^2= 2 (k_2^2 + k_3^2 + \vec{k}_2\cdot\vec{k}_3)\, ,
\ee
is symmetric under permutations and Lorentz invariant.

The spin-flavor functions are given explicitly by sums over the following 
products of Clebsch-Gordan coefficients:
\bea
\chi_{\ohalf,\sigma,\tau}(\sigma_1,\tau_1,\sigma_2,\tau_2,\sigma_3,\tau_3)&:=&
 \sqrt{\half}\Bigl\{ \delta_{\sigma,\sigma_1}
(\half,\half\sigma_2,\sigma_3|0,0)(\half,\half\tau_2,\tau_3|0,0)
\cr
&+& (\half,\half\sigma_2,\sigma_3|1,\sigma_2+\sigma_3)
(1,\half,\sigma_2+\sigma_3,\sigma_1|\half, \sigma)\cr
&\times&
(\half,\half\tau_2,\tau_3|1,\tau_2+\tau_3)
(1,\half,\tau_2+\tau_3,\tau_|\half, \tau)\Bigr\}\; ,
\label{SPINFLAV}
\eea
and
\bea
\chi_{\thalf,\sigma,\tau}(\sigma_1,\tau_1,\sigma_2,\tau_2,\sigma_3,\tau_3)&:=&
(\half,\half\sigma_2,\sigma_3|1,\sigma_2+\sigma_3)
(1,\half,\sigma_2+\sigma_3,\sigma_1|\thalf, \sigma)\cr
&\times&
(\half,\half\tau_2,\tau_3|1,\tau_2+\tau_3)
(1,\half,\tau_2+\tau_3,\tau_1 |\thalf, \tau)\; .
\eea
The spatial part of the wave function is parameterized by functions
that depend only on the hyperspherical momentum variable, which
is defined as
$\P:= \sqrt{2 (\vec\kappa^2+\vec q^2)}$:
\begin{equation}
\varphi_{0}(\P)= {\cal N} 
\left( 1+ {\P^2\over 4 b^2}\right)^{-a},
\label{ground}
\end{equation}
where ${\cal N}$ is a normalization constant and the exponent $a$ and
$b$ are adjustable parameters.

It was noted in Ref.~\cite{boffi} that by introduction of a 
small admixture of a mixed symmetry $S-$state component in the nucleon
neutron electric form factors that agree with extant data may be 
obtained. For that purpose a mixed symmetry $S-$wave component is 
also considered here. Its detailed construction is described 
in Appendix~\ref{app:neutron}.

The radial wave function for the $N(1440)$ is constructed so that it 
is orthogonal to the ground state, and that its Fourier transform, 
$\tilde \varphi_1({\sf R})$, has a single node:
\be
\tilde \varphi_1(\vec \rho,\vec r):={1\over (2\pi)^3}\int d^3\kappa\int d^3 q
e^{-\imath(\vec \kappa\cdot \vec \rho+\vec q\cdot \vec r)}
\varphi(\vec \kappa,\vec q) \,.
\ee

These conditions imply the following general form in the momentum representation:
\beq
\varphi_{1}(\P) = A\varphi_0(\P)+
b^2 \;B\left[{20\over \P} \varphi_0'(\P) +4\varphi_0''(\P)\right] \,,
\label{roper}
\eeq
where $A$ and $B$ are parameters, which are determined by
the orthonormality condition 
\beq
 \int d^3{\kappa} \; d^3{q} \; \varphi_i(\vec{\kappa}, \vec{q})
\varphi_j(\vec{\kappa}, \vec{q}) = \delta_{ij}. 
\label{norm1}
\eeq

Given the ground state wave function model, (\ref{ground}), the explicit
expression for the wave function $\varphi_1(\P)$ is 
\begin{equation}
\varphi_1(\P) = \varphi_0(\P) \left(A(a) 
+B(a)\left[ -12a {1\over (1+{\P^2\over 4 b^2})}
+ {a(a+1) \over b^2}  {\P^2 \over (1+{\P^2\over 4 b^2})^2} \right]\right).
\label{roperexpl}
\end{equation} 

The rms radius of the quark distribution of the nucleon is given by the 
expression
\be
r_0^2 =\thalf \int d^3 \rho\int d^3 r \rho^2 \; | \tilde{\varphi}({\sf R})|^2
\; , \qquad{ \sf R}:=\sqrt {2(\rho^2+r^2)}\; .
\ee
It follows that $r_0 \sim 1/b$.
        
\begin{figure}[t]
\vspace{15pt}
\includegraphics[width=8cm]{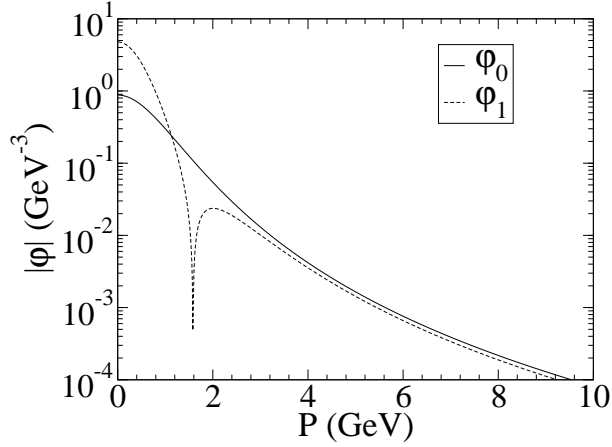}
\caption{Ground state and first radially excited 
states as a function of 
${\sf P}= \sqrt{2(\kappa^2+q^2)}$ for $b=640$ MeV and $a=9/4$.\label{fig:wf}}
\end{figure}

In Fig.~\ref{fig:wf} $\varphi_0$ and $\varphi_1$ are shown for the parameter 
value $b=640$ MeV and $a=9/4$. In Table~\ref{parameters} the values of the 
two parameters $a$ and $b$ used in the following sections for the different 
forms of kinematics are listed along with the corresponding values of the 
quark radius $r_0$.

\begin{table}[b]
\caption{Values of the parameters of the ground 
state wave function used for the three different 
forms of kinematics. The corresponding matter radii
$r_0$ are listed in the last column.\label{parameters}}
\begin{ruledtabular}
\begin{tabular}{lcccc}
       & $m_q$ (MeV)  & $b$ (MeV) & $a$ &$r_0$ (fm) \\ 
\hline
point form   &  350           &  640        &  9/4 & 0.19 \\
front form   &  250           &  500        &   4  & 0.55 \\ 
instant form &  140       &  600        &   6  & 0.63  \\
\end{tabular}
\end{ruledtabular}
\end{table}
\subsection{ Quark currents}
For each form of kinematics the dynamics generates the current-density operator 
from a kinematic current, which is specified by the expression:
\be
\bra{\vec v_f,\vec v_2',\vec v_3'}I^\mu(0)
\ket{\vec v_3,\vec v_2,\vec v_a}=\delta^{(3)}(v_3'-v_3)
\delta^{(3)}(v_2'-v_2)\left(\osix+\half \tau_3^{(1)}\right)\bar u(\vec v_1\,')
\gamma^{(1)\mu}u(\vec v_1)\, ,
\label{cur4}
\ee
in the case of Lorentz kinematics, and:
\bea
&&\bra{P^+, P_{\perp f},{\bf p}_2',{\bf p_3}'}I^+(x^-,x_\perp)
 \ket{{\bf p}_3,{\bf p_2}, P_{\perp a},P^+}\cr
&&=\delta^{(3)}(p_3'-p_3)
\delta^{(3)}(p_2'-p_2)(\osix+\half \tau_3^{(1)})
\bar u({\bf p_1'}) \gamma^{(1)+}u({\bf p_1})
e^{\imath( P_{\perp f}- P_{\perp a})\cdot  x_\perp}\, ,
\eea
for light-front kinematics and finally by:
\bea
&&\bra{\half \vec  Q,\vec p'_2,\vec p_3'}I^\mu(\vec x)
 \ket{\vec p_3,\vec p_2,-\half \vec Q }=\delta^{(3)}(p_3'-p_3)
\delta^{(3)}(p_2'-p_2)(\osix+\half \tau_3^{(1)})\nonumber\\
&&\bar u(\vec p_1\,')\gamma^{(1)\mu}u(\vec p_1)
e^{\imath(\vec Q\cdot \vec x)}\, ,
\eea
for instant kinematics. In each case only covariance under the kinematic 
subgroup is required.

For the corresponding expression for the axial vector current
the matrix is obtained by the replacement:
\be
\left(\osix+\half \tau_3^{(1)}\right)\gamma^{(1)\mu}\;\to\;
\left(\gamma^{(1)}_\mu 
+ g_P{(p_1'-p_1)^\mu\over 2m}\right) \gamma^{(1)}_5 \ohalf\tau^{(1)}_3\, . 
\label{axialc}
\ee
The value for the pseudoscalar coupling constant of 
the ``partially conserved'' axial current is then
\beq
g_P = {4 m^2\over Q^2 + m_\pi^2}\, .
\label{pseudosc}
\eeq

\subsection{Velocity representations of the quark structure}

With Lorentz kinematics quark momenta $p_i:=m v_i$ are defined by the 
boost relations:
\be
p_i:= B(v)\{0, \vec k_i\}+ \omega_i v_K\;\, , \qquad 
\omega_i:=\sqrt{|\vec k_i|^2+m^2} \; , \qquad v_K^2=-1\;\,.
\label{point1}
\ee
The Lorentz covariance of the quark velocity operators follows from the 
Lorentz covariance (\ref {VEL}) and (\ref{KWIG}) of the operator $v$ and the
momenta $\vec k_i$. 
The quark velocities do not commute with any component of the 
momentum $P=\M v$, and the sum of the quark momenta does not equal the 
total momentum for any component.
 
Null-plane kinematics depends on a null vector $n$ or the ``frame'' in which
\newline $n=\{-1,0,0,1\}$. The quark momenta $ p_i$ defined by
\be
p_i^+= \xi_i P^+\; , \qquad 
p_{i\perp}:= k_{i\perp}+ \xi_i P_\perp\; \qquad p_i^-:={m^2+p_{i\perp}^2\over p_i^+}
\qquad\mbox{with}\qquad \xi_i:= {k_{iz}+\omega_i\over \sum_i \omega_i}\,,
\label{PNULL}
\ee
 are covariant only under the subgroup which leaves the 
null-plane $n\cdot x= 0$ invariant.

With instant form kinematics the kinematic subgroup does not include 
any boosts. The kinematic symmetry of the quark momenta merely requires 
covariance under rotations and $\sum_i \vec p_i= \vec P$. That much allows 
considerable freedom in the relation of the quark momenta $\vec p_i$ to 
the internal momenta. In Ref.~\cite{coe98} the momenta $\vec p_i$ were defined 
as functions of the boosted Jacobi momenta $B_c(v) \{0,\vec \kappa\}$ , $B_c(v) \{0,\vec q\} $ 
and the total momentum $\vec P$. That definition had the virtue of formal 
simplicity. Here the velocities $\vec p_i/m$ are taken to
be free-quark 
velocities, which implies
\bea
p_i= \omega_i v_K+ B_c(v_K) \{0,\vec k_i\}\ &=& 
\{v_{Kz}k_{iz}+\omega_iv_K^0,k_{ix},k_{iy},v_K^0k_{iz}+\omega_i v_{Kz}\}\ ,\cr
\vec k_i &=& \{p_{ix},p_{iy},v_K^0p_{iz}-E_i v_{Kz}\}\ \, ,
\label{INPK}
\eea
with $\vec v_K:= \vec P/\sum_i\omega_i$. Canonical boosts are used because 
then the Wigner rotations of rotations are identical to the rotations. 

The quark momenta so defined  by either Eq. (\ref{PNULL}) or  (\ref{INPK})
are covariant under the 
kinematic subgroup, that is 
\be 
U^\dagger(\Lambda)p_iU(\Lambda)= \Lambda p_i\, ,
\ee
with $\Lambda$ restricted to the kinematic subgroup.

For the three forms of kinematics changes in the representation 
of the baryon states:
\bea
\{\vec \kappa ,\vec q,\vec v\}&\to& \{\vec p_2,\vec p_3,\vec v\}\qquad
\mbox {with}\qquad
\vec p_1 =\vec p_1(\vec p_2, \vec p_3,\vec v)\, , \nonumber\\
\{\vec \kappa ,\vec q ,{\bf  P}\} &\to& \{{\bf p_2},{\bf p_3},{\bf P}\}
\qquad\mbox {with}\qquad
{\bf p_1}={\bf P}-{\bf p_2} - {\bf p_3} \, ,\nonumber \\
\{\vec \kappa ,\vec q ,\vec P\} &\to& \{\vec p_2,\vec p_3,\vec P\}
\qquad\mbox {with}\qquad
\vec p_1=\vec P-\vec p_2-\vec p_3 \, , 
\eea
together with the corresponding spin to spinor transformations~\cite{coe92},
\bea
u_c(v_i)&=&{\vec \alpha\cdot \vec v_i+\beta +v_i^0
\over \sqrt{2v_i^0(1+v_i^0)}}{1+\beta\over2}\, ,\\
u_f(\bf p_i)&=& {\vec\alpha_\perp \cdot\vec p_\perp+\beta m +p^+
 \over
\sqrt{2 m p^+}}{1+\alpha_3\over 2} {1+\beta \over 2} \, ,\\
u_c(\vec p_i)&=&{\vec \alpha\cdot \vec p_i+m\beta +E_i
\over \sqrt{2E_i(m+E_i)}}{1+\beta\over2}\; ,
\qquad  E_i:=\sqrt{m^2+|\vec p_i|^2} \, ,
\eea 
provide kinematically covariant representations of the baryon states, which
are convenient for the construction of conserved current density operators
which satisfy Poincar\'e covariance.

In each case the wave function must be multiplied by the square 
root of the appropriate Jacobian. For Lorentz kinematics this is
\be
J(\vec v;\vec p_2,\vec p_3):=
\left({\partial (\vec \kappa,\vec q) \over 
\partial (\vec p_2,\vec p_3)}\right)_{\vec v}
= {\sqrt{27} \omega_2 \omega_3\over E_2E_3}=
 {\sqrt{27}(E_2 v^0-p_{2z}v_z)(E_3 v^0-p_{3z}v_z)\over E_2E_3}\, .
\label{pointjac}
\ee
With null plane kinematics the Jacobian is
\bea
J(\bf P;\bf p_2,\bf p_3)&:=&
\left({\partial( \vec \kappa,\vec q) \over \partial ({\bf  p}_2, {\bf p}_3)} \right)_{\bf P}=
{\partial(\vec \kappa,\vec q)\over \partial(\xi_2,k_{2\perp},\xi_3,k_{3\perp})}
\,\left({ \partial(\xi_2,k_{2\perp},\xi_3,k_{3\perp})\over
\partial( {\bf  p_2},{\bf p_3})}\right)_{\bf P}\cr\cr
&=&{\sqrt{27}\omega_1\omega_2\omega\,_3 (P^+) \over p_1^+p_2^+p_3^+ 
(\omega_1+\omega_2+\omega_3)}\, ,
\eea
with
\be
\omega_i=\half\left(\xi_i M_0+ {m^2+k_{i\perp}^2\over \xi_i M_0}\right)\; , \qquad
M_0^2= \sum_i{m^2+k_{i\perp}^2\over \xi_i}= \left(\sum_i \omega_i\right)^2\, .
\ee
With instant kinematics the definition (\ref{INPK}) implies the Jacobian:
\newpage
\bea
J(\vec P,\vec p_2,\vec p_3)&:=& 
\left({\partial \vec \kappa,\vec q)\over \partial (\vec p_2,\vec p_3)}\right)_{\vec P}
=\sqrt{27} {\partial(\vec k_2,\vec k_3,\vec v) \over \partial(\vec p_2,\vec p_3,\vec v)}\;
  {\partial(\vec k_2,\vec k_3,\vec P) \over \partial(\vec k_2,\vec k_3,\vec v)}\;
 {\partial(\vec p_2,\vec p_3,\vec v) \over \partial(\vec p_2,\vec p_3,\vec P)}\cr\cr
 &=& {\sqrt{27} \omega_2 \omega _3 \over E_2 E_3} \left( 1- v_z {k_{1z}\over E_1}\right)
\eea
where
\be
P_x=P_y=0\; , \qquad M_0^2= (\sum _i E_i)^2-|\vec P|^2 \; , \qquad  \vec v := {\vec P\over M_0} \,.
\ee

With canonical boosts the spin variables must be transformed by the required momentum
dependent rotation matrices 
\be
D^\half_{\lambda_i,\sigma_i}\left(R_W[B(v_K),k_i]\right) \qquad \mbox {with}
\qquad
R_W[B(v_K),k_i]:= B^{-1}(p_i)B(v_K)B(k_i)\; .
\ee

With canonical boosts explicit representations of these 
Wigner rotations are,
\bea
D^{1/2} \Bigl({\cal R}_W[B(v_{K}),k_i] \Bigr)= \cos{\theta_i\over 2}-
\imath \sin{\theta_i\over 2}\,
{(\vec p_i\times \vec \sigma_i)_z\over |p_{i\perp}|} \; , \qquad i= 1,2,3 \, , 
\label{RWIG1}
\eea
where the angles $\theta_i$ are defined by
\be
\sin{\theta_i\over2}= -{v_{Kz} |p_{i\perp}|
\over\sqrt{ 2(1+v_{K}^0)(m+E_i)(m+\omega_i)}} \,.\label{THETA}
\ee.

With null-plane kinematics the corresponding required spin rotations are 
Melosh rotations, which are represented by~\cite{coe92}
\beq
D^\half[\R_M(\vec k _i)]
= { m+\xi_i M_0 -\imath \vec{\sigma} \cdot (\vec{n} \times \vec{k}_{i\perp}) \over
         [(m+\xi_i M_0)^2 + \vec k_{i\perp}^2 ]^{1/2} } \, .
\label{RMEL}
\eeq

\section{Nucleon form factors}

\subsection{Canonical spin representations }

The matrix elements (\ref{GEGM}) may be evaluated with the antisymmetric 
nucleon wave function (\ref{ground}) and the quark current (\ref{cur4}) 
multiplied by 3 (the number of constituent quarks). Evaluation of the sum 
over spin and isospin indices leads to the explicit 
expressions of the form factors 
\begin{eqnarray}
G_E(\eta)&=&\int d^3 p_2 d^3 p_3 \,
\varphi\left({{\kappa'}^2+q^{'2}\over 2 b^2}\right)
\varphi\left({\kappa^2+q^2\over 2 b^2}\right)
\sqrt{{\cal J}_{fa}(\vec p_2,\vec p_3)} 
C_{23}(\eta,\vec p_2,\vec p_3){\cal S}_e(\eta,\vec p_2,\vec p_3),
\cr\cr\cr
G_M(\eta)&=&\int d^3 p_2 d^3 p_3 
\varphi\left({{\kappa'}^2+q^{'2}\over 2 b^2}\right)
\varphi\left({\kappa^2+q^2\over 2 b^2}\right)
\sqrt{{\cal J}_{fa}(\vec p_2,\vec p_3)}
C_{23}(\eta,\vec p_2,\vec p_3){\cal S}_m(\eta,\vec p_2,\vec p_3)\; .\cr
&&
\label{GEM}
\end{eqnarray}
The Jacobian factor ${\cal J}_{fa}$,
\beq
{\cal J}_{fa} := J(v_f,\vec p_2,\vec p_3)
J(v_a,\vec p_2,\vec p_3)\, ,
\label{JAC}
\eeq
is defined by Eq.~(\ref{pointjac}) or~(\ref{instjac}) for Lorentz or 
instant kinematics respectively.

The coefficient $C_{23}(\eta,\vec p_2,\vec p_3)$ is determined by 
the spectator Wigner rotations:
\bea
C_{23}(\eta,\vec p_2,\vec p_3)&=&{1\over 2} \sum_{\sigma',\sigma}
\left[\sum _{ \sigma''}
{\D^{1/2}_{\sigma' \sigma''}}^\dagger\left({\cal R}_W[B(v_{Kf}), k'_2]\right)
\D^{1/2}_{\sigma'',\sigma}\left({\cal R}_W[B(v_{Ka}), k_2]\right)\right]\cr\cr
&\times&
\left[ \sum_{\sigma''}{\D^{1/2}_{-\sigma' \sigma''}}^\dagger\left({\cal R}_W[B(v_{Kf}), k'_3]\right)
D^{1/2}_{\sigma'',-\sigma}\left({\cal R}_W[B(v_{Ka}), k_3]\right)\right]\; .
\label{C23}
\eea
The velocities $\vec v_{Ka}$, $\vec v_{Kf}$ are $\vec v_a$, $\vec v_f$ 
with Lorentz kinematics and $\half \vec Q/\M_0'$, $-\half \vec Q/\M_0$ 
with instant kinematics.

The factors ${\cal S}_e$ and ${\cal S}_m$ arise from the Dirac spinor 
structure of the current and associated Wigner rotations. The explicit 
expressions are
\begin{eqnarray}
{\cal S}_e&=&\bra{\half }{D^\half}^\dagger[\R_W(v_{Kf},k_1')] \,
\bar u_c(v_1') \, \gamma^0 \, u_c(v_1) \, 
D^\half[\R_W(v_{Ka},k_1)]\ket{\half}\cr\cr
&=&\sqrt{{(E_1'+m)(E_1+m)(1+\eta)\over 4 E_1' E_1}}
\bigg\{\left[1+ {\vec p_1\,'\cdot \vec p_1 \over 
(E_1'+m)(E_1+m)} \right] \cos\left({\theta_1-\theta_1'\over 2}\right)
\cr\cr && 
+{|p_{1\perp}|(p_{1z}' -p_{1z}) \over
 (E_1'+m)(E_1+m)}\sin\left({\theta_1-\theta_1'\over 2}\right)  \bigg \}\;,
\label{Se}
\end{eqnarray}
and
\begin{eqnarray}
{\cal S}_m&= &{1\over \sqrt{\eta}}
\bra{\half}{D^\half}^\dagger[\R_W(v_{Kf},k_1')]\,\bar u_c(v_1')
\,\gamma_x \,u_c(v_1) \,D^\half[\R_W(v_{Ka},k_1)]\, \ket{-\half}\cr\cr
&=& \sqrt{{1+\eta \over  4\eta E_1' (E_1'+m) E_1 (E_1+m)}}
\bigg\{
\Bigl[ p_{1z}'(E_1+m )- p_{1z} (E'_1+m) \Bigr]
\cos{\theta_1'\over2}\cos{\theta_1\over 2}\cr\cr
&+&{1\over2}|p_{1\perp}|(E_1'+E_1+2m)\sin{\theta_1'-\theta_1\over2}
+{1\over2}|p_{1\perp}|(E_1-E_1')\sin{\theta_1'+\theta_1\over2}\bigg\},  
\label{Sm}
\end{eqnarray}
respectively. The boost dependent angles of rotation of the initial and
final spins of the struck constituent are defined in Eq.(\ref{THETA}).

In Appendix~\ref{app:deltacan} and~\ref{app:axialinst} the corresponding 
expressions for the matrix elements relevant for the calculation of the axial 
and $N\to\Delta$ transition form factors are given.

\subsection{Light-front spin representations}

The form factors  of the proton, $\tau=\half$, and the neutron, $\tau=-\half$,
\be
F_{1,\tau}(Q^2) := \bra{ \ohalf,\tau} I^+(0) \ket{\tau,\ohalf}\, , \qquad
F_{2,\tau}(Q^2): = {1\over \sqrt{\eta}} \bra{-\ohalf,\tau} I^+(0) \ket{\tau,\ohalf}, 
\ee
can be written 
in a compact form as,
\beqa
F_{\alpha,\tau}(Q^2) &=& \int_0^1 d\xi_1 \,d\xi_2 \,d\xi_3 \, {\delta(\xi_1+\xi_2+\xi_3-1)
 \over\xi_1\xi_2\xi_3}\,\nonumber \\
&& \int d\vec{k}_{1\perp}\, d\vec{k}_{2\perp}\, d\vec{k}_{3\perp}\, 
\delta(\vec{k}_{1\perp}+\vec{k}_{2\perp}+\vec{k}_{3\perp})
 \sqrt{{ \omega_1\omega_2\omega_3\over M_0}
 {\omega_1'\omega_2'\omega_3'\over M_0'}}\, \nonumber \\
&&
\varphi^*(\xi_1,\vec{k}_{1\perp}',\xi_2,\vec{k}_{2\perp}',\xi_3,\vec{k}_{3\perp}')
{\cal F}_{\alpha,\tau}(Q^2)
\varphi(\xi_1,\vec{k}_{1\perp},\xi_2,\vec{k}_{2\perp},\xi_3,\vec{k}_{3\perp})\; ,
\eeqa
where 
\be
k_{1\perp}'= k_{1\perp}+(1-\xi_1)Q_\perp\; , \qquad 
k_{i\perp}'= k_{i\perp}-\xi_1Q_\perp \qquad i=2,3\; .
\ee

The factors ${\cal F}_{\alpha,\tau}(Q^2)$ involve the spin-isospin amplitudes
(\ref{SPINFLAV}) and  the effects of the Melosh 
rotations on both spectators and the quark current $I_{q1}$, 
\beqa
{\cal F}_{1,\tau}(Q^2) &=& \chi_{\half,\tau}^\dagger
\, D^{1/2\dagger}({\cal R}_{M1'}) \, 
\,I^+_{q_1} \, 
D^{1/2}({\cal R}_{M1}) \; \nonumber \\ &\times& 
D^{1/2\dagger}({\cal R}_{M2'}) 
D^{1/2}({\cal R}_{M2}) 
D^{1/2\dagger}({\cal R}_{M3'})
D^{1/2}({\cal R}_{M3}) \,
\chi_{-\half,\tau}\, , \nonumber \\
{\cal F}_{2,\tau}(Q^2)
 &=&{1\over \sqrt{\eta}} \chi_{-\half,\tau}^\dagger
\, D^{1/2\dagger}({\cal R}_{M1'}) \, 
\,I^+_{q_1} \, 
D^{1/2}({\cal R}_{M1}) \;\nonumber \\ &\times&  
D^{1/2\dagger}({\cal R}_{M2'}) 
D^{1/2}({\cal R}_{M2}) 
D^{1/2\dagger}({\cal R}_{M3'})
D^{1/2}({\cal R}_{M3}) \,
\chi_{\half,\tau}\, .
\eeqa
 The representations of the Melosh rotations are defined in Eq.~(\ref{RMEL}).

In the explicit evaluation of the integrals the choice $p_{2\perp}=p_{2x}$
is made without loss of generality. This leads to the following 
expressions for ${\cal F}_i(Q^2)$ :
\beqa
{\cal F}_{1p}&=& {1\over D_1 D_2 D_3} f_1 [ f_2 f_3 +V_{2y}V_{3y}]\, , \nonumber \\
{\cal F}_{2p}&=& {2m_p\over Q} {1\over D_1 D_2 D_3} (-V_{1y}) [ f_2 f_3 +V_{2y}V_{3y}]
\, , \nonumber \\
{\cal F}_{1n}&=& {-1\over3}{1\over D_1 D_2 D_3} [2f_1 V_{2y}V_{3y} -f_2 V_{1z}V_{3z}
-f_2V_{1x}V_{3x}-f_2 V_{1y}V_{3y} -f_3 V_{1y}V_{2y}] \, , \nonumber \\
{\cal F}_{2n}&=& {2\over3}{m_p\over Q} {1\over D_1 D_2 D_3} 
[2V_{1y} f_2f_3 -V_{1x}V_{2y}V_{3x}-f_1f_2V_{3y}-f_1f_3V_{2y} 
-V_{1z}V_{3z}V_{2y}]\, .
\eeqa
Here the following notation has been employed:
\beqa
a_i  &=&  m +\xi_i M_0\, ,  \qquad
a_i' =  m +\xi_i M_0'\, ,  \qquad \vec{n}= (0,0,1)\, ,  \nonumber \\
D_i  &=&  \sqrt{ {a'}_k^2+ \vec{k'}_{i\perp}^2 } \;
          \sqrt{ a_k^2+ \vec{k}_{i\perp}^2 }\, , \nonumber \\
f_i  &=&  a'_i a_i + \vec{k'}_{i\perp} \cdot \vec{k}_{i\perp}\, , \nonumber \\
\vec{V}_k &=& -a_k (\vec{n} \times \vec{k'}_{i\perp}) +
               a'_k (\vec{n} \times \vec{k}_{i\perp}) +
                     (\vec{n}\times \vec{k'}_{i\perp})
 \times  (\vec{n}\times\vec{k}_{i\perp})\, .
\label{frontnot}
\eeqa     

In Appendix~\ref{app:af} the corresponding 
expressions for the matrix elements relevant for the 
computation of the axial form factors are given.

\section{Numerical results}
\subsection{Nucleons}
\subsubsection{Finite values of the constituent mass}

In order to explore qualitative differences of the three forms of kinematics 
nucleon form factors were calculated with the mass and current operators 
specified in Sec. III with the parameter values listed in Table~\ref{parameters}.

\begin{figure}[bt]
\vspace{25pt}
\begin{center}
\mbox{\epsfig{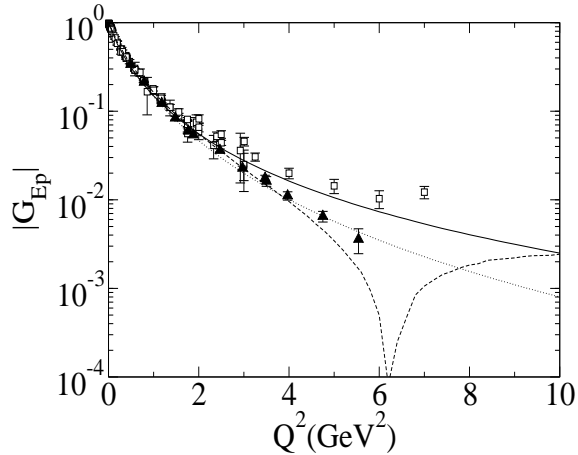}}
\end{center}
\caption{Electric form factor of the proton. 
Solid, dotted and dashed lines correspond to 
the instant, point and front forms respectively. 
Squares are from the compilation of 
Ref.~\protect\cite{gep} while black triangles are obtained
from the recent JLAB data of Refs.~\protect\cite{jones} 
using \\$G_{Ep} = (\mu_p G_{Ep} / G_{Mp})/(1+Q^2/0.71)^2$.\label{fig:gep}}
\end{figure}

\begin{figure}[bt]
\vspace{25pt}
\begin{center}
\mbox{\epsfig{file=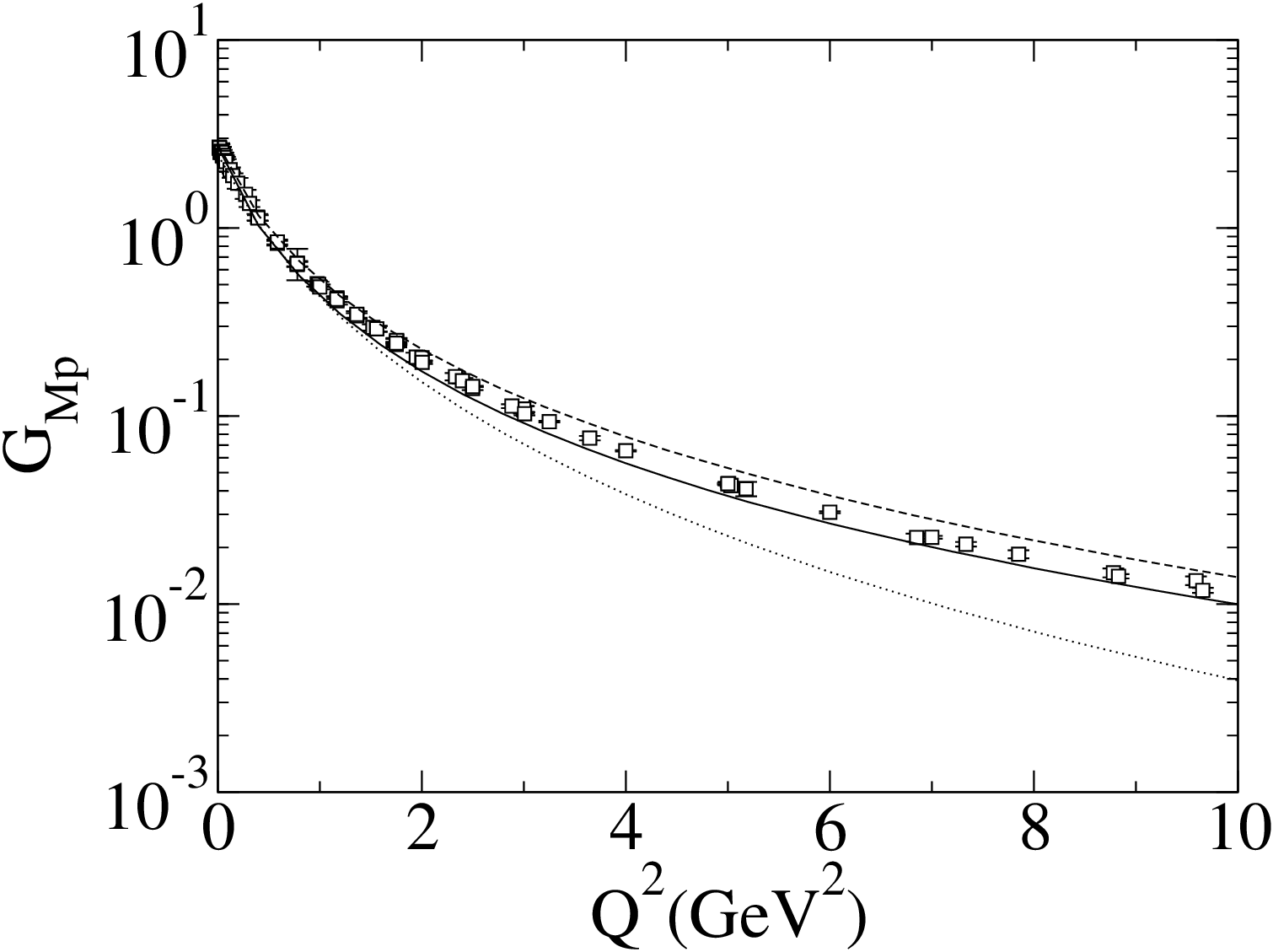, width=75mm, height=60mm}}
\end{center}
\caption{Magnetic form factor of the proton. 
Solid, dotted and dashed lines correspond to 
the instant, point and front forms respectively. 
The experimental data are from the compilation of 
Ref.~\protect\cite{gep}.\label{fig:gmp}}
\end{figure}

\begin{figure}[p]
\vspace{25pt}
\begin{center}
\mbox{\epsfig{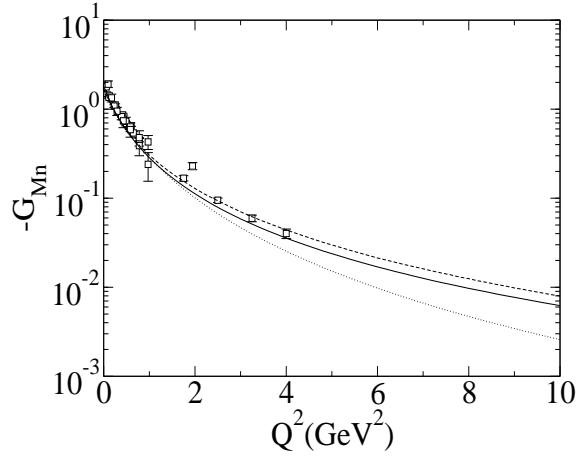}}
\end{center}
\caption{Magnetic form factors of the neutron. 
Solid, dotted and dashed lines correspond to 
the instant, point and front forms respectively. 
The experimental data are from the compilation of 
Ref.~\protect\cite{gep}.\label{fig:gmn}}
\end{figure}

\begin{figure}[p]
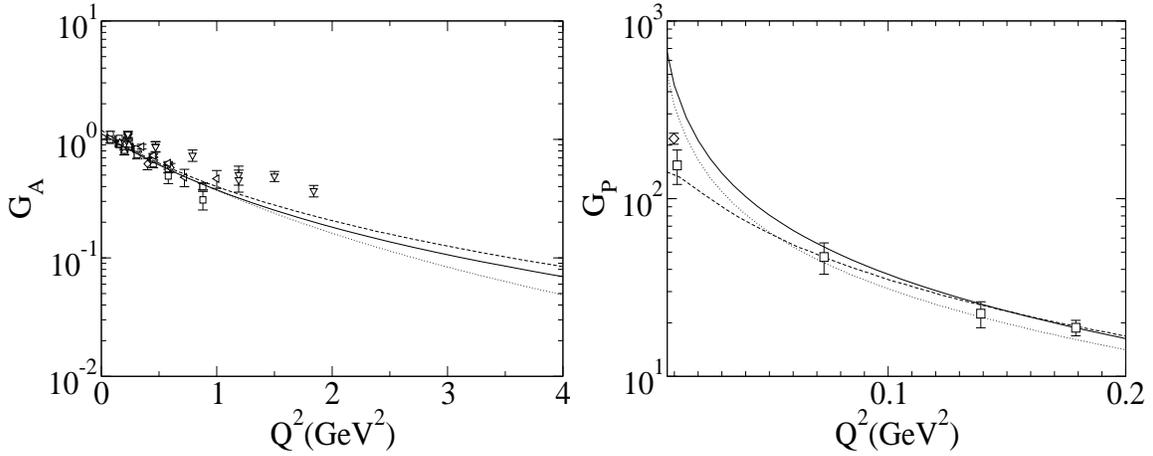

\vspace{25pt}
\begin{center}
\mbox{\epsfig{file=fig6a, width=75mm, height=60mm}}
\mbox{\epsfig{file=fig6b, width=75mm, height=60mm}}
\end{center}
\caption{Axial and induced pseudoscalar
form factors of the nucleon. 
Solid, dotted and dashed lines correspond to 
the instant, point and front forms respectively. 
The experimental data are from 
Refs.~\protect\cite{ga,gp}.\label{fig:ga}}
\end{figure}

The calculated form factors are shown in 
Figs.~\ref{fig:gep},~\ref{fig:gmp},~\ref{fig:gmn} and~\ref{fig:ga} 
along with data taken from the compilation~\cite{gep}.
In Table~\ref{zeroq} the corresponding values of $G_{Mp}(0)$, $G_{Mn}(0)$, 
$G_A$(0) together with the proton charge radii are listed.
The rms radius $r_0$ of the wave function is always smaller than the
charge radius with the largest value for instant kinematics and 
the smallest for Lorentz kinematics.

\begin{table}[b]
\caption{Values of the form factors at $Q^2\rightarrow 0$ together
with the proton charge radius.\label{zeroq}}
\begin{ruledtabular}
\begin{tabular}{lcccc}
                &  Instant &  Point  &  Front & EXP \\  
\hline
 $G_{Mp}(0)$    &   2.7    &  2.5   & 2.8   & 2.793~\protect\cite{PDG}  \\
 $G_{Mn}(0)$    &   $-$1.8   &  $-$1.6  & $-$1.7  & $-$1.913~\protect\cite{PDG}\\
 $G_A(0)$       &   1.1    &  1.1   & 1.2   &  1.2670~\protect\cite{PDG}\\ 
 $r_{cp}$(fm)   &  0.89   & 0.84  & 0.85  & 0.87~\protect\cite{PDG}\\
\end{tabular}
\end{ruledtabular}
\end{table}

The results reveal that it is possible to reach agreement
with the empirical data for all these form factors. 
The electric form factor of the proton is found to be more 
sensitive to the form of kinematics than the magnetic form 
factors. That may be related to the implementation 
of current conservation, which involves the form of dynamics.

\begin{figure}[bt]
\vspace{25pt}
\begin{center}
\mbox{\epsfig{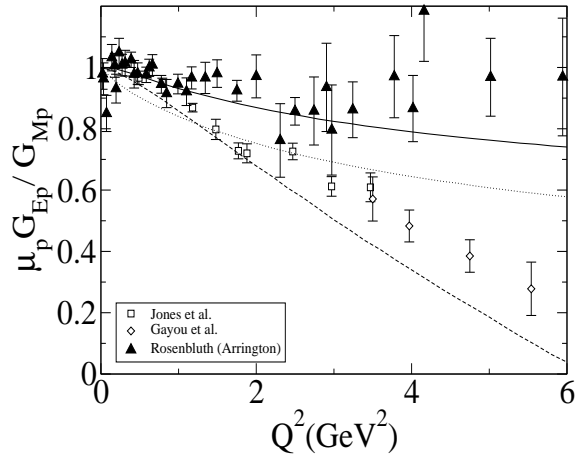}}
\end{center}
\caption{Quotient $\mu_p G_{Ep}/ G_{Mp}$ compared 
to the recent experimental data measured in 
TJNAF, Refs.~\protect\cite{jones,arrington}. Solid, dotted and dashed 
lines correspond to the instant, point and front forms 
respectively. 
\label{fig:gepogmp}}
\end{figure}

The instant form result for $G_{Ep}$ in Fig.~\ref{fig:gep} follows 
the empirical values obtained by a Rosenbluth separation up to 6 GeV$^2$. 
The point form calculation, with a very compact wave function, 
follows the data that have been obtained by means of polarization 
transfer somewhat more closely. The front form calculation of the 
form factors $F_1$ and $F_2$ produces cancellations of $F_1$ and 
$\eta F_2$ at about $6$ GeV$^2$ \cite{Miller,brodsky}. Such behavior 
is in fact suggested by the recent experimental data for the quotient 
$\mu_p G_{Ep}/ G_{Mp}$~\cite{jones}. In Fig.~\ref{fig:gepogmp} this 
ratio is shown as calculated with the three forms of kinematics. This 
figure emphasizes the differences between the three forms of kinematics as 
well as the discrepancies in the form factors of the proton at 
medium energies. This discrepancy between the recent TJNAF data,
measured using the polarization transfer technique, 
and the previous data, obtained through the Rosenbluth 
separation~\cite{arrington}, could be partly due to two photon 
exchanges as recently explored in Ref.~\cite{blunden}. 
There is therefore a qualitative difference between models based 
on canonical-spin representations of the currents (instant and Lorentz 
kinematics) and models based on null-plane-spin representations of the 
currents (front-form kinematics).

Similar results for the elastic form factors have been described in 
the literature making use of point~\cite{boffi,coe03} and front~\cite{ck95} 
forms of kinematics on the basis of different dynamical models.

The calculated values of the magnetic form factors 
shown in Figs.~\ref{fig:gmp} and~\ref{fig:gmn} are in qualitative 
agreement with each other and the data. The calculated magnetic moments 
show significant differences.
With instant form kinematics reasonable agreement with the empirical 
values of the nucleon magnetic moments requires a very small quark 
mass of 140 MeV. With larger quark mass values the magnitude of the 
calculated magnetic moments is too small. This feature also appears
in the non-relativistic quark model with ``relativistic corrections''~\cite{dannb}. 
The missing strength is in that model attributed to exchange current contributions.

The magnetic moment values that are obtained in point form 
kinematics are about 10 \% too small. This feature was already 
noted in Ref.~\cite{boffi}. In this case the calculated values
of the magnetic moments are fairly insensitive to the quark mass value.

The magnetic moment of the proton as calculated in front form kinematics
with a wave function of intermediate range also falls within 1 \%
of the empirical value. In front form kinematics the 
calculated neutron magnetic moment falls some 12 \% below the
corresponding empirical value.

The calculated value of the axial vector coupling constant, $G_A(0)$, is
closest to the empirical value (within 6 \%) when front form kinematics is
used. In instant and point form kinematics the calculated value is about
14 \% smaller than the empirical value. These values differ significantly 
from the static quark model value $5/3$, which is too large by 31 \%.

The calculated values of the axial form factor are close to the
empirical values \cite{ga} in all forms of kinematics as shown in 
Fig~\ref{fig:ga}. The pseudoscalar form factor follows the empirical 
values, except at very small values of momentum transfer, where only 
the front-form one actually goes through the muon point. 

\subsubsection{The electric form factor of the neutron}

\begin{figure}[b]
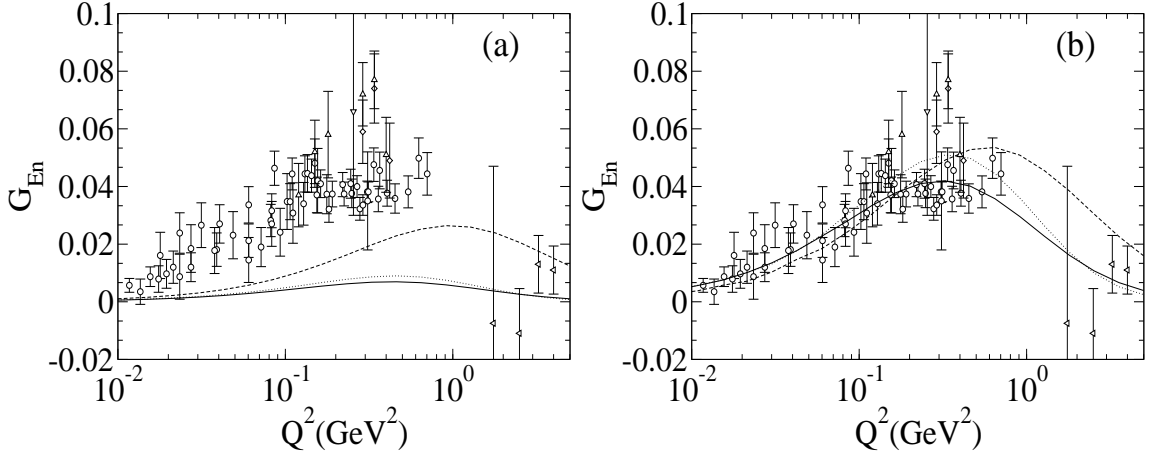

\vspace{20pt}
\begin{center}
\mbox{\epsfig{file=fig7a, width=75mm, height=60mm}}
\mbox{\epsfig{file=fig7b, width=75mm, height=60mm}}
\end{center}
\caption{Electric form factor of the neutron. Solid, dotted and
dashed lines correspond to the instant, point and front forms respectively.
(a) No mixed symmetry $S-$state is included.
(b) Some percentage of mixed symmetry $S-$state
is included in the neutron wave function as described in the text. 
Experimental data are from Refs.~\protect\cite{gep}.\label{GEN}}
\end{figure}

Without the momentum dependent spin rotations (\ref{RWIG1}) or
(\ref{RMEL}) the symmetric spin-isospin amplitude (\ref{SPINFLAV})
would imply a vanishing electric form factor of the neutron.
Because of Wigner rotations or Melosh rotations of the constituent 
spins the electric form factor of the neutron does not vanish.
This is shown in Fig.~\ref{GEN}a. The magnitude is however negligible
with instant and point form kinematics and too small with front form 
kinematics in agreement with the results of Refs.~\cite{Chung, Simula}.

In Ref.~\cite{boffi}, using point form kinematics, it was noted that a small 
admixture of a mixed symmetry $S-$state in the ground state
may produce a satisfactory form factor. The effects of a $2\%$ admixture 
of a mixed symmetry $S-$state wave function are also shown in Fig.~\ref{GEN}b
for all three forms of kinematics. The results are in good agreement
with the empirical data. The agreement is not quite as good with 
a 1\% admixture of the mixed symmetry $S-$state and would deteriorate 
with a larger admixture. 

The mixed symmetry $S-$state is represented by appropriate
combination of mixed symmetry spin-isospin wave functions with
two radial wave functions of mixed symmetry of the form
\begin{equation}
\varphi_s (\vec \kappa,\vec q)= {\cal N}_s{\kappa^2-q^2\over\kappa^2
+q^2}\varphi_0(\kappa,q)\, ,\quad
\varphi_a (\vec \kappa,\vec q)= {\cal N}_a{\vec\kappa\cdot\vec q\over\kappa^2
+q^2}\varphi_0(\kappa,q)\, .
\label{mix}
\end{equation}
where $\varphi_0(\vec\kappa,q)$ is the symmetric $S-$state wave 
function (\ref{ground}). The explicit construction is given in 
Appendix~\ref{app:neutron}.

\begin{figure}[pbt]
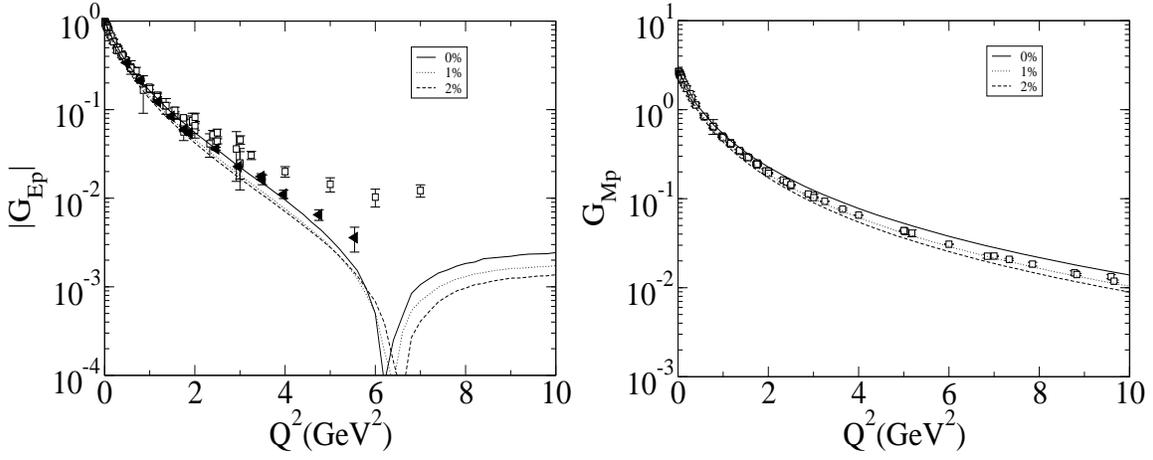

\vspace{20pt}
\begin{center}
\mbox{\epsfig{file=fig8a, width=75mm, height=60mm}}
\mbox{\epsfig{file=fig8b, width=75mm, height=60mm}}
\end{center}
\caption{Electric (left) and magnetic (right) form 
factors of the proton calculated with front-form 
kinematics with some mixed symmetry $S-$state component as 
explained in the text. Solid, dotted and
dashed lines correspond to $0\%$, $1\%$ and $2\%$ of mixed 
symmetry $S-$state. Squares are from the compilation of 
Ref.~\protect\cite{gep} while black triangles are obtained
from the recent JLAB data of Refs.~\protect\cite{jones} 
using $G_{Ep} = (\mu_p G_{Ep} / G_{Mp})/(1+Q^2/0.71)^2$.\label{GEMPmix}}
\end{figure}

The effect of the introduction of the mixed symmetry $S-$state component 
on the other nucleon form factors is small. This is illustrated
in Fig.{\ref{GEMPmix}}, where the modification of the calculated front 
form electric and magnetic proton form factors by the mixed symmetry 
$S-$state is shown. The values of the slope $(dG_{en}/ dQ^2)_{Q^2=0}$ that 
are obtained are 0.60 GeV$^{-2}$, 
0.56 GeV$^{-2}$ and 0.39 GeV$^{-2}$ for instant, point 
and front form respectively, while 
the experimental value is 0.511$\pm$0.008 GeV$^{-2}$ 
\cite{genexp}.

\clearpage
\subsubsection{The zero quark mass limit}

It has been noted~\cite{coe03} that in the case of point form 
kinematics and spectator currents the form factors were insensitive 
to unitary scale transformations of the wave functions when
the extent of the wave function was small compared to 
the scale defined by the quark mass, $<r^2> \ll 1/m_q^2$. 
This is equivalent to $b^2 \gg m_q^2$. In the ``point'' limit $ m_q^2/b^2 =0$ 
the calculated form factors are invariant under unitary scale 
transformations. This opens the possibility for quark model phenomenology 
with a very small constituent mass. 

For the class of quark models considered here, where the representation 
of the baryon mass operator is independent of the quark mass, a zero-quark-mass 
limit of the form factors exists for all three forms of kinematics.
The spinor representations of the quark currents and the boost transformations
to spin representations are also independent of the quark mass.
Only the Jacobians and the Wigner or Melosh rotations depend
on the quark mass.
 
\begin{figure}[h]
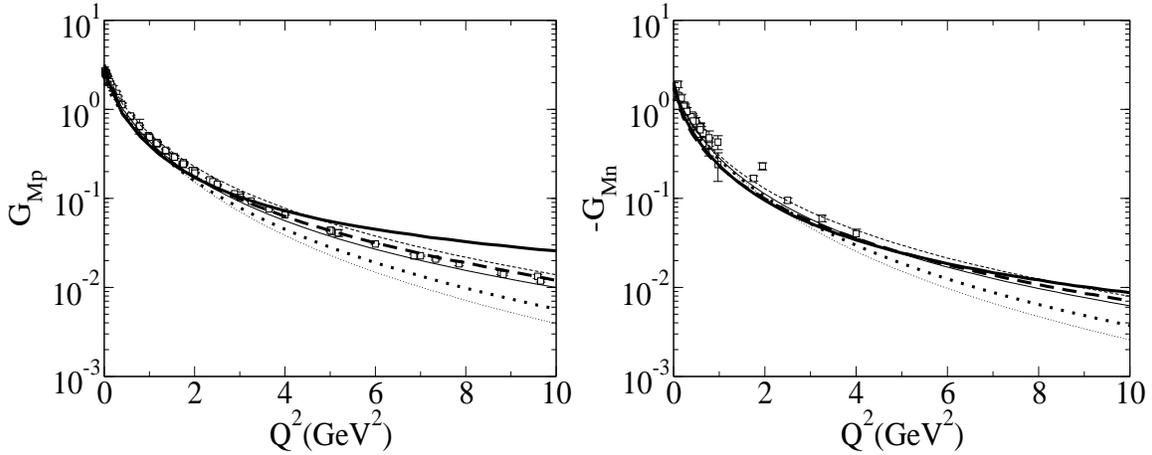

\vspace{20pt}
\begin{center}
\mbox{\epsfig{file=fig9a, width=75mm, height=60mm}}
\mbox{\epsfig{file=fig9b, width=75mm, height=60mm}}
\end{center}
\caption{Magnetic form factor of the proton(left), 
of the neutron(right). Solid, dotted and dashed lines 
correspond to the instant, 
point and front forms respectively. 
Thick lines correspond to the zero quark mass case. Experimental 
data are from Refs.~\protect\cite{gep}.
\label{fig:pointy1}}
\end{figure}

\begin{figure}[p]
\vspace{10pt}
\begin{center}
\mbox{\epsfig{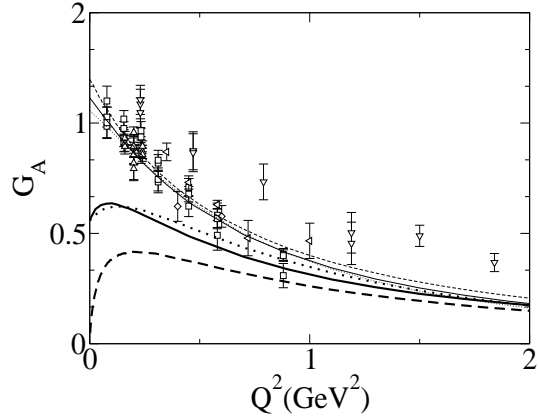}}
\end{center}
\caption{Axial form factor. Solid, dotted and dashed lines 
correspond to the instant, 
point and front forms respectively. 
Thick lines correspond to the zero quark mass case. Experimental 
data are from Refs.~\protect\cite{ga}.
\label{fig:ga0}}
\end{figure}

Zero-quark-mass values of the magnetic moments and the axial coupling 
constant are listed in Table~\ref{zerom}. In Figs.~\ref{fig:pointy1} 
and~\ref{fig:ga0} the zero-mass limits of nucleon magnetic and axial form 
factors are compared to the finite-mass values. 
The magnetic moments of both the protons and the neutrons 
do not show large changes with vanishing mass in the case of 
instant form kinematics.
The results for the nucleon magnetic form factors, see Fig.~\ref{fig:pointy1}, 
in the zero mass limit show that the form factors are insensitive 
to the quark mass only with Lorentz kinematics.

\begin{table}[b]
\caption{Values of the form factors at $Q^2\rightarrow 0$
for zero constituent mass.~\label{zerom}}
\begin{ruledtabular}
\begin{tabular}{lcccc}
                &  Instant  &  Point  &  Front & EXP \\  
\hline
 $G_{Mp}(0)$    &   2.6     &  2.   & 3.2  & 2.793~\protect\cite{PDG}  \\
 $G_{Mn}(0)$    &   $-$1.7    & $-$1.3 & $-$2.0 & $-$1.913~\protect\cite{PDG}\\
 $G_A(0)$       &   0.6     &  0.6  & 0.0  &  1.2670~\protect\cite{PDG}\\ 
\end{tabular}
\end{ruledtabular}
\end{table}

\begin{figure}[p]
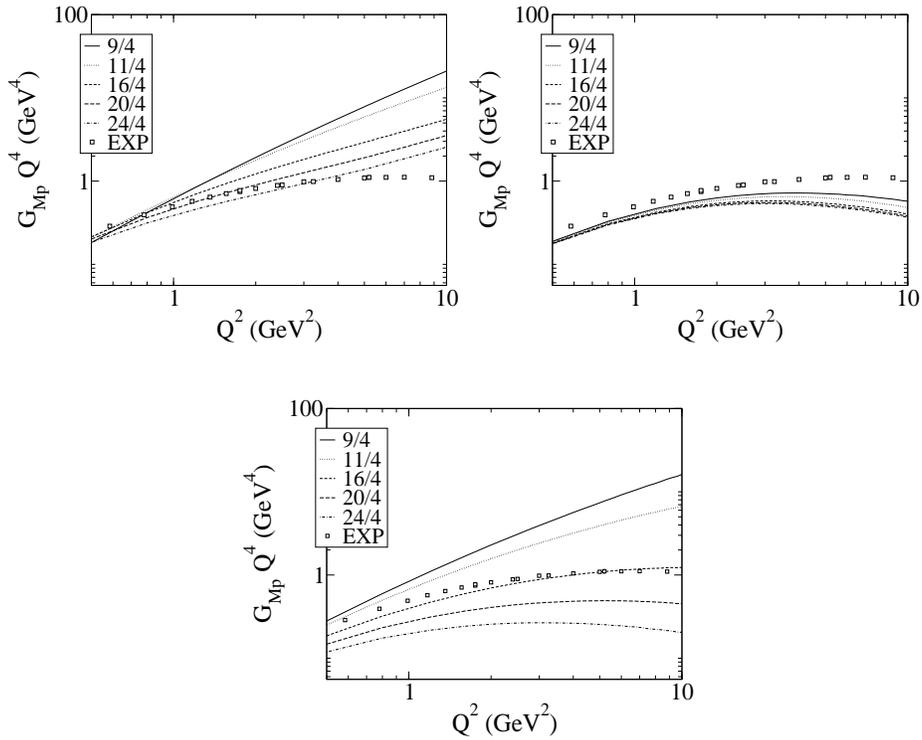

\vspace{20pt}
\begin{center}
\mbox{\epsfig{file=fig11a, width=60mm, height=45mm}}
\mbox{\epsfig{file=fig11b, width=60mm, height=45mm}}
\vspace{20pt}

\mbox{\epsfig{file=fig11c, width=60mm, height=45mm}}
\end{center}
\caption{Magnetic form factor of the proton for $m_q=0$ in 
instant(above-left), point(above-right) and front(below) forms 
respectively. The different lines correspond to different 
values of the exponent $a$ in Eq.~(\protect\ref{ground}).
Experimental data are from Refs.~\protect\cite{gep}.
\label{fig:pointy2}}
\end{figure}

Fig.~\ref{fig:pointy2} shows magnetic 
form factors computed with zero quark mass and wave functions with different 
rms radii obtained varying the exponent $a$ with fixed $b$. Thus both the range
and the shape of the wave function are changed. The point form results
show the expected  scale independence and indicate
that the change in shape is relatively unimportant.
With instant and front-form kinematics. Fig.~\ref{fig:pointy2} shows the expected
drastic changes   in the $Q^2$ dependence 
In particular, instant form 
calculations do not reproduce the experimental 
behavior at high-$Q^2$ for any of the exponents, while
front form ones do give the correct behavior for the 
appropriate value of $a$.

The zero mass limit does not yield satisfactory
values for the axial coupling constant and the
axial form factor, as seen in Table \ref{zerom} and
in Fig.\ref{fig:ga0}. These results suggest that realistic axial 
current phenomenology in the constituent quark model demands 
that the constituent mass at least be of the order of 200 MeV.
Overall, the zero-quark-mass limit is not a good candidate for 
quark-model phenomenology if both axial and electromagnetic 
properties are to be understood simultaneously.

\clearpage
\subsection{The $\Delta(1232)\rightarrow N$ transition form factors}

\begin{figure}[htb]
\vspace{20pt}
\begin{center}
\mbox{\epsfig{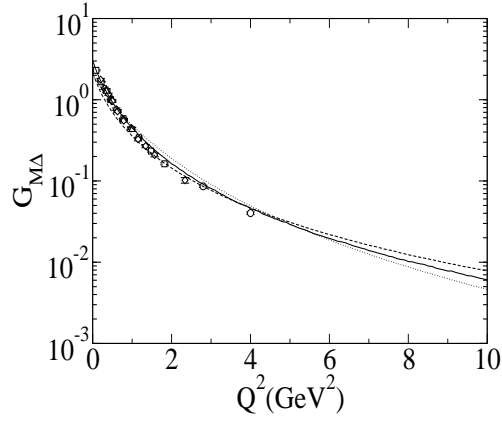}}
\end{center}
\caption{Magnetic $\Delta \rightarrow N$ transition form factor. 
Solid, dotted and dashed lines correspond to the instant, point and front 
forms respectively. Experimental data are 
from Refs.~\protect\cite{gnd}.\label{fig:gnd}}
\end{figure}

The magnetic transition form factor $G_{M\Delta}$ that is associated with
the $\Delta(1232)-N$ transition as calculated with the wave 
function (\ref{ground}) and the parameter values in Table \ref{parameters}
in the three forms of kinematics is shown in Fig.~\ref{fig:gnd}. The 
corresponding values for the transition magnetic moments are listed 
in Table~\ref{DeltaN}.

In the case of instant form kinematics the impulse approximation
describes the empirical form factor and the transition magnetic
moment well. This is a notable improvement compared to 
non-relativistic quark models. The magnetic moment 
is too small by about 30 \% in both point and front 
kinematics. 

\begin{table}[b]
\caption{The $\Delta(1232)-N$ transition magnetic moment in
the different forms of kinematics with finite
and zero constituent mass.~\label{DeltaN}}
\begin{ruledtabular}
\begin{tabular}{lcccc}
                            &  Instant  &  Point  &  Front & EXP \\  
\hline
 $G_{M\Delta}(0)$ $m\neq 0$ &   2.8     &  2.0    & 2.1  & 3.1  \\
 $G_{M\Delta}(0)$ $m=0$     &   3.0     &  1.7    & 2.5  & 3.1  \\
\end{tabular}
\end{ruledtabular}
\end{table}

\begin{figure}[hbt]
\vspace{20pt}
\begin{center}
\mbox{\epsfig{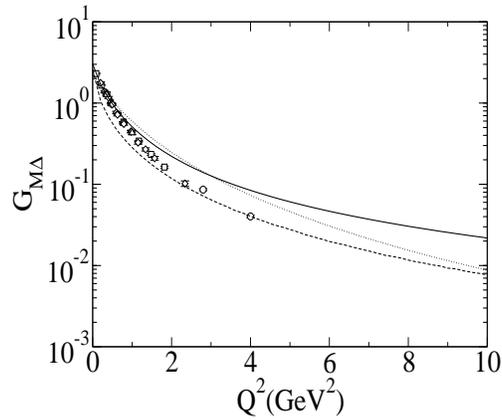}}
\end{center}
\caption{Magnetic $\Delta \rightarrow N$ transition form factor in the 
zero quark mass case. Solid, dotted and dashed lines correspond to the 
instant, point and front forms respectively. Experimental data are from 
Refs.~\protect\cite{gnd}.\label{fig:gnd0}}
\end{figure}

In Table~\ref{DeltaN} the transition magnetic moments are also
listed as obtained in the zero quark mass case. The corresponding 
form factors are plotted in Fig.~\ref{fig:gnd0}. These results
are fairly similar to those obtained with finite values
of the constituent mass.

\subsection{The $N(1440)\rightarrow N$ form factors}

\begin{figure}[hpbt]
\vspace{20pt}
\begin{center}
\mbox{\epsfig{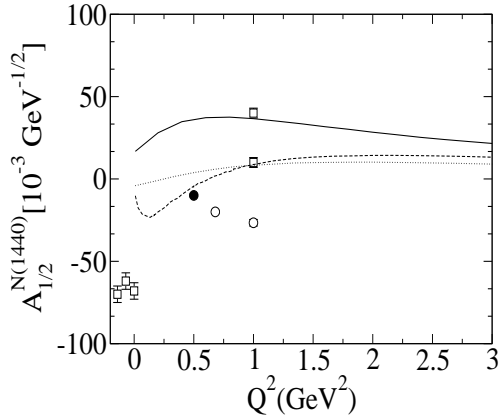}}
\end{center}
\caption{Helicity amplitude for $N^*(1440)$ electroexcitation. Solid, dotted and 
dashed lines correspond to the instant, point and front forms respectively. 
The experimental data are from Refs.~\protect\cite{exproper}.
The full solid point also corresponds to a preliminary analysis from CLAS
~\protect\cite{CLAS}.
\label{fig:a12}}
\end{figure}

In Fig.~\ref{fig:a12} we show the calculated helicity
amplitude $A_{1/2}$, defined as:
\beq
A_{1/2} =\sqrt{{ 4 \pi \alpha \over 2 E_\gamma}} \sqrt{\eta} G_M 
\qquad E_\gamma := {M^{*2} -M^2 \over 2 M^*} \quad \alpha \sim 1/137 \,,
\eeq
for the $N(1440)-N$ transition as
obtained with the wave function models~(\ref{ground})
and~(\ref{roperexpl}) in all forms of kinematics with the
parameter values in Table~\ref{parameters}. 
It is, of course, questionable whether a treatment of the
$N(1440)$ as a stable 3-quark bound state is realistic. 
The extant data of this helicity
amplitude shown in Fig.~\ref{fig:a12} 
\cite{exproper,CLAS} are manifestly inadequate.
The calculated transition form factors do in this case depend
significantly on the form of kinematics. 

\begin{figure}[hpbt]
\vspace{20pt}
\begin{center}
\mbox{\epsfig{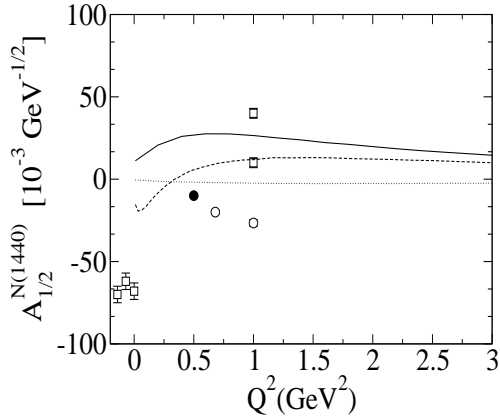}}
\end{center}
\caption{Helicity amplitude for $N^*(1440)$ electroexcitation in the 
zero quark mass case. 
Solid, dotted and dashed lines correspond to the instant, point and
front forms respectively. 
The experimental data are from Refs.~\protect\cite{exproper}.
The full solid point also corresponds to a preliminary analysis from CLAS
~\protect\cite{CLAS}.
\label{fig:a120}}
\end{figure}

In Fig~\ref{fig:a120} the helicity amplitude obtained
with zero quark mass is shown. These results are qualitatively 
similar to those that are obtained with finite values of the 
constituent mass above. 

\clearpage
\newpage
\section{Conclusions}

The present comparative study of how baryon form factors
calculated with free-quark currents depend on the choice 
of the kinematic subgroup revealed a number of features,
which might be of phenomenological utility in description 
of the baryons by constituent-quark models.
This exploration is based on mass operators of confined 
quark, which are parameterized by a confinement scale, and 
which implement basic symmetries without free-quark features 
or dependence on quark masses. Poincar\'e covariant current 
density operators are generated by the dynamics from 
free-quark current densities covariant under different kinematic 
subgroups 

The interpretation of baryon wave functions of constituent-quark models
as a description of a physical structure that is observed by electro-weak
processes depends on the choice of a form of kinematics.
To assess the effectiveness of a choice of kinematics it is important to 
consider the full range of elastic and inelastic transitions at low and 
medium energies. 
Examination of a broad range of features with a crude 
model structure revealed no drastic failure that would rule out
any of the forms of kinematics considered. For most form factors permutation
symmetric $S-$wave functions were adequate. The electric form factor of the
neutron required a small mixed-symmetry admixture. 
The baryon wave functions
used for this exploration are independent of the quark mass
and dependent on a range parameter and a shape parameter. 
Significantly different values of these parameters are required
for adequate wave functions with different forms of kinematics.
A quantitative phenomenology for all available form factor data would 
require different refinements with different forms of kinematics. 

The calculated form factors are functions of kinematic quantities, which 
differ with form of kinematics and the baryon masses. Since at least one
component of the four-momentum is dynamic, current conservation will always
imply some dependence on the baryon masses.

The features emphasized by instant form kinematics are closest to 
those of non-relativistic quark models with a physical interpretation of
the wave function that emphasizes covariance under 3-dimensional rotations and 
translations. Lorentz boosts are not in the kinematic subgroup. There is no
kinematic Lorentz symmetry of quark velocities. The kinematic variable of
the form factors is $\vec Q^2$ which equals the four-momentum transfer 
only for elastic transitions.

With light-front kinematics both relevant Lorentz boosts and translations are 
kinematic. The corresponding subgroup is Galilean 
symmetry in $1+2$ dimensions with $p_i^+$ in the role of the masses. 
In that case rotations about the direction of the momentum transfer are not 
kinematic. In this the kinematic parameter of the form factors is the 
four-momentum transfer $Q^2$

With point-form kinematics there are no kinematic translations and there
is no kinematic interpretation of the wave function as a representation of 
spatial structure. The kinematic parameter of the form factors is the 
invariant velocity transfer $\eta$.

\begin{acknowledgments}

The authors want to thank Jun He, Robert F. Wagenbrunn and S. Simula for 
helpful correspondence concerning earlier versions of this work.
B. J.-D. thanks Dirk Merten for providing many of the 
experimental data. B. J.-D. thanks the European Euridice network for 
support (HPRN-CT-2002-00311). Research supported in part by 
the Academy of Finland through grant 54038 and by the U.S. Department 
of Energy, Nuclear Physics Division, contract W-31-109-ENG-38.

\end{acknowledgments}

\clearpage
\appendix

\section{Invariant form factors for $N\to \Delta$ transition}
\label{appnd}

The best way to relate calculations made in the three 
forms of kinematics to the standard invariant form factors
is through the following set of invariant form factors, 
similar to those used in Ref.~\cite{Scad},
\beq
\Gamma^\mu_\nu (P,Q)= \sum_i^3 {\cal K}_\nu^{i,\mu}(P,Q)G_i(Q^2)\, ,
\eeq
with,
\beqa
 {\cal K}^{1,\nu\mu}(P,Q) &=& {Q^\nu \gamma^\mu - (\gamma\cdot Q) 
g^{\nu\mu}\over \sqrt{Q^2}}\sqrt{M^* M} 
\gamma_5\, , \nonumber \\
 {\cal K}^{2,\nu\mu}(P,Q) &=& {Q^\nu P^\mu - (P\cdot Q) g^{\nu\mu
}\over \sqrt{Q^2}} \;\gamma_5\, , \nonumber \\
 {\cal K}^{3,\nu\mu}(P,Q) &=& {Q^\nu Q^\mu - Q^2
 g^{\nu\mu}\over Q^2 } M^* \;\gamma_5 \, ,
\eeqa
where $M$ and $M^*$ are the nucleon and resonance masses respectively.
The Sachs like magnetic dipole, electric quadrupole and Coulomb 
form factors are defined as in Ref.~\cite{Scad},
\beq
\Gamma^\mu_\nu (P,Q)= 
G_M^*(Q^2) {\cal K}_\nu^{M,\mu}(P,Q)
+G_E^*(Q^2) {\cal K}_\nu^{E,\mu}(P,Q)
+G_C^*(Q^2) {\cal K}_\nu^{C,\mu}(P,Q) \,.
\eeq
The relation between the two sets of form factors is 
given by,
\beqa
G_E^* &=&{M\over 3(\MD+M)} 
\bigg[
{\MD^2-M^2-Q^2\over \MD}\sqrt{\MD M \over Q^2} G_1
+{\MD^2-M^2\over \sqrt{Q^2}}G_2
-2 \MD G_3 
\bigg]\, ,\label{EMC} \\
G_M^* &=&{M\over 3(\MD+M)}
\bigg[ 
{(3\MD+M)(\MD+M)+Q^2\over \MD}\sqrt{\MD M \over Q^2}G_1 
+{\MD^2-M^2\over \sqrt{Q^2} } G_2 
- 2 \MD G_3\bigg]\, ,\nonumber \\
G_C^* &=& {2M\over 3(\MD+M)} 
\bigg[ 2 \MD\sqrt{\MD M \over Q^2} G_1 
+{3\MD^2+M^2+Q^2 \over 2 \sqrt{Q^2}} G_2
+{\MD^2-M^2-Q^2\over Q^2} \MD G_3\bigg] \,.  
\nonumber 
\eeqa

The relation between matrix elements 
in the different forms of kinematics and the form factors $G_i$
is provided below.

\subsection{Canonical representation}

\begin{eqnarray}
\langle \thalf, P_\Delta\vert I_1(0)\vert P_N,\ohalf \rangle
&=&\left[ {M^* +M\over \sqrt{ Q^2}} G_1 + 
{M^{*2}-M^2\over 2\sqrt{Q^2 M M^*}}G_2
-{\sqrt{M^*\over M} }G_3\right]
  {Q_3\over 2 \sqrt{E(M+E)}} \, ,
\nonumber\\
\langle \ohalf, P_\Delta\vert I_1(0)\vert P_N,-\ohalf\rangle
&=&-{\sqrt{3}\over 6}\left[{M^*+M\over \sqrt{ Q^2}}G_1
+ {M^{*2}-M^2\over 2\sqrt{ Q^2 M M^*}}G_2-
{\sqrt{M^*\over M}}G_3\right]
  {Q_3\over \sqrt{E(M+E)}} \, ,
\nonumber \\
&+&{\sqrt{3}\over 3}{Q_3\over \sqrt{Q^2}}{M+E\over 
\sqrt{E(M+E)}}G_1\, ,
\label{Gs} \\
\langle \ohalf, P_\Delta\vert I_0(0)\vert P_N,\ohalf\rangle
&=&- {\sqrt{3}\over 3}
\left[ {Q_3\over \sqrt{Q^2}}G_1
+{Q_3\over \sqrt{Q^2M M^*}}{E+M^*\over 2} G_2
+{Q_3 Q_0 \over Q^2}\sqrt{M^*\over M} G_3 \right] 
 {Q_3 \over \sqrt{ E (M+E)} } \, . \nonumber
\end{eqnarray}
where,
\beq
Q=\{Q^0,0,0,Q_3\} \qquad Q^0 = -{P^* \cdot Q \over M^*} =  {M^{*2} - M^2 -Q^2\over 2 M^*}; 
\qquad  Q_3=\sqrt{Q^2+Q^{02}} \, .
\eeq

\subsection{Light-Front representation}

\beqa
\bra{\thalf} I^+(0) \ket{\ohalf} &=& 
-{1\over \sqrt{2}} \left[G_1+ {\MD -M \over \sqrt{4 M\MD}}G_2 \right] \,,
\\
\bra{\ohalf} I^+(0) \ket{-\ohalf} &=&{1\over \sqrt{6}} 
\biggl[ 
-{M\over \MD} G_1 
+ { (\MD-M) -Q^2/\MD \over  \sqrt{4M\MD}} G_2
+\sqrt {Q^2\over M\MD} \,G_3 
\biggr] \,,
\nonumber \\
\bra{\ohalf} I^+(0) \ket{\ohalf}&=& {1\over \sqrt{6}} \biggl[ 
 {\sqrt{Q^2} \over \MD}G_1
+\sqrt{{Q^2\over 4\MD M}} \left({\MD-M \over \MD}+1\right) G_2 
-{\MD-M \over \sqrt{M\MD}} G_3 
\biggr]\,. 
\nonumber 
\eeqa

\section{Neutron mixed symmetry $S-$state}
\label{app:neutron}
Consider the following two components in the neutron wave 
function,
\beq
\ket{n} =  A \, \ket{n,S} + B\, \ket{n,M_S} \, ,
\eeq
where,
\beq
\langle n,S \vert n,S\rangle =\langle n,M_S \vert n,M_S \rangle=1\, ,
 \qquad
\langle n,S \vert n,M_S\rangle =\langle n,M_S \vert n,S \rangle =0\, ,
\qquad |A|^2+|B|^2=1 \, .
\eeq
The two components are in more explicit form,
\beqa
\ket{n,S}   &=& [3]_x \, [3]_{FS,S}\, , \nonumber \\
\ket{n,M_S} &=& {1\over \sqrt{2}} \left( [21]_{x,S} [21]_{FS,S} +
 [21]_{x,A} [21]_{FS,A} \right) \,,
\eeqa
where 
the spatial part has been indicated by an $x$, the explicit 
expression of which is,
\beqa
[3]_x      &=&\varphi_G(P)= \varphi(P)\, , \nonumber \\
\left[21\right]_{x,S} &=&\varphi_S(P)=
 {\cal N}_S\, { \kappa^2 - q^2 \over\kappa^2 + q^2} \, \varphi(P)\, , \nonumber \\
\left[21\right]_{x,A} &=&\varphi_A(P)=
 {\cal N}_A\, {\vec{\kappa}\cdot\vec{q}\over \kappa^2 + q^2} \, \varphi(P) \, ,
\eeqa
where $P=\sqrt{2( \kappa^2 +q^2)}$ with
 ${\cal N}_S$ and ${\cal N}_A$ obtained normalizing the spatial wave functions
as,
\beq
{1\over \sqrt{27}} \int d^3 \,q d^3\kappa \;
 \varphi_i^* \varphi_j = \delta_{ij} \, .
\eeq

The Flavor-Spin wave functions can be written as,
\beqa
\left[3\right]_{FS,S} &=& {1\over 
\sqrt{2}} \left( [21]_{F,S} [21]_{S,S} + 
[21]_{F,A} [21]_{S,A} \right)\, , \nonumber \\
\left[21\right]_{FS,S} &=& {1\over \sqrt{2}} 
\left( [21]_{F,S} [21]_{S,S} - [21]_{F,A} [21]_{S,A} \right)\, , \nonumber \\
\left[21\right]_{FS,A} &=& {1\over \sqrt{2}} 
\left( [21]_{F,S} [21]_{S,A} + [21]_{F,A} [21]_{S,S} \right) \,
\eeqa
in terms of spin and flavor wave functions of three quarks.

\subsection{General Matrix Element}
A general matrix element between two neutron states can be 
written,
\beq
\bra{n} \,{\cal X\,Q\,S }\, , \ket{n}
\eeq
where ${\cal X}$ is an spatial operator, ${\cal Q}$ is a flavor operator, the charge
in this case, and ${\cal S}$ is an spin operator.
This expression can be worked out to arrive 
to the following general expression:
\beqa
\bra{n}\, {\cal X\,Q\,S} \, \ket{n}  &=&
[  A^* \, \bra{n,S} + B^*\, \bra{n,M_S}] 
\,{\cal  X\,Q\,S }\, 
[  A \, \ket{n,S} + B\, \ket{n,M_S}]\, , \nonumber \\
&=&
|A|^2   \bra{n,S}\,{\cal  X\,Q\,S} \, \ket{n,S} \nonumber \\
&+&
A^*B    \bra{n,S}\, {\cal X\,Q\,S} \,  \ket{n,M_S}\nonumber \\
&+&
B^*A    \bra{n,M_S}\, {\cal X\,Q\,S} \,  \ket{n,S}\nonumber \\
&+&
|B|^2    \bra{n,M_S}\, {\cal X\,Q\,S} \, \ket{n,M_S}\, ,
\eeqa
with
\beq
\bra{n,S}\,{\cal  X\,Q\,S} \, \ket{n,S} =
{1\over 6} [3]_x \,{\cal  X} \, [3]_x \; 
\bigg[ [21]_{S,S} \,{\cal S} \,  [21]_{S,S}  -  
[21]_{S,A}\,{\cal S} \,  [21]_{S,A} \bigg] \, ,
\eeq

\beqa
\bra{n,M_S}\, {\cal X\,Q\,S} \, \ket{n,M_S} 
&=& 
{1\over 12} \bigg(
\bigg[  [21]_{S,S} \,{\cal S}\, [21]_{S,S}  -
  [21]_{S,A}  \,{\cal S}\, [21]_{S,A} \bigg] \nonumber \\
&&\bigg[  [21]_{x,S} \,{\cal X} \, [21]_{x,S}  - 
 [21]_{x,A} \, {\cal X}\, [21]_{x,A} \bigg] \nonumber \\
&&+
\bigg[   [21]_{S,A} \,{\cal S}\, [21]_{S,S} + 
 [21]_{S,S} \,{\cal S}\,[21]_{S,A} \bigg] \nonumber \\
&&\bigg[  [21]_{x,A}  \,{\cal X} \, [21]_{x,S}+
 [21]_{x,S}\, {\cal X} \, [21]_{x,A} \bigg]\bigg)  \, ,
\eeqa

\beqa
\bra{n,S}\, {\cal X\,Q\,S} \,  \ket{n,M_S} &=&
{1\over 6\sqrt{2}} \bigg( 
[3]_x \,  {\cal X}\, [21]_{x,S}\;
\bigg[  [21]_{S,S} \,{\cal S}\, [21]_{S,S}  +  [21]_{S,A}  \,{\cal S}\, [21]_{S,A} \bigg]\nonumber \\
&&+
[3]_x \,  {\cal X}\,  [21]_{x,A} \;
\bigg[   [21]_{S,S}\,{\cal S}\, [21]_{S,A} -  [21]_{S,A}\,{\cal S}\, [21]_{S,S}  \bigg] \bigg) \, ,
\eeqa
and,
\beqa
\bra{n,M_S}\, {\cal X\,Q\,S} \,  \ket{n,S} &=&
{1\over 6 \sqrt{2}} \bigg(
[21]_{x,S}\,  {\cal X}\,[3]_x \,  
\bigg[  [21]_{S,S} \,{\cal S}\, [21]_{S,S}  +  [21]_{S,A}  \,{\cal S}\, [21]_{S,A} \bigg]
\nonumber \\
&+&
[21]_{x,A} \,  {\cal X}\,[3]_x \, 
\bigg[   [21]_{S,A} \,{\cal S}\, [21]_{S,S} -  [21]_{S,S} \,{\cal S}\,[21]_{S,A} \bigg]
\bigg) \,.
\eeqa

\section{$\Delta \to N$ matrix elements in canonical representation}
\label{app:deltacan}

The relevant combination entering in the calculation of 
the magnetic form factor takes the explicit form:
\beqa
&& \bra{\ohalf, \ohalf}I_+\ket{\thalf, -\ohalf} +
\sqrt{3} \bra{\ohalf, -\ohalf}I_+ \ket{\thalf, -\thalf} \, =\,
 {8\over 3\sqrt{2}}
{ 1\over 2\sqrt{E_1'E_1(E_1'+m)(E_1+m)}}
\nonumber \\
&&\left\{ 
[c_2c_3 
-{s_2s_3\over 4}\hat{p}_{2\perp}\cdot \hat{p}_{3\perp} ]
\left (A \cos{\beta'\over2} \cos{\beta\over2} 
+B \sin{\left({\beta'+\beta\over2}\right)}
+C \sin{\left({\beta'-\beta\over2}\right)} \right )\right.
\nonumber \\
&&+ 
{1\over4}s_2 s_3\;\hat{P}_{13}\hat{P}_{12}
\left(
A\sin{\beta'\over2} \sin{\beta\over2}
-B\sin{\left({\beta'+\beta\over2}\right)}
+C\sin{\left({\beta'-\beta\over2}\right)} \right)
\nonumber \\
&&+ 
\left[ c_2 s_3  \hat{P}_{13}
+ c_3 s_2  \hat{P}_{12}  \right]
\left(
-A{1\over8} \cos{\beta'\over2} \sin{\beta\over2}
-A{5\over8}\sin{\beta'\over2}\cos{\beta\over2} \right.\nonumber \\
&&+\left.\left.
B{3\over4}
\cos{\left({\beta'+\beta\over2}\right)}
+{1\over2}C\cos{\left({\beta'-\beta\over2}\right)}
\right) \right\}\, ,
\eeqa
with
\beqa
&&A=[ p_{1z}' (E_1+m) -p_{1z} (E_1'+m)] \, , \quad
B=|p_{\perp}| (E_1-E_1')\, , \quad
C= |p_{\perp}|(E_1+E_1'+2m)\, ,  \nonumber \\
&&\hat{P}_{12}={(p_{1x}+ip_{1y})\over |p_{\perp}|} 
{(p_{2x}-ip_{2y})\over |p_{2\perp}|} \, , \qquad
\hat{P}_{13}={(p_{1x}+ip_{1y})\over |p_{\perp}|} 
{(p_{3x}-ip_{3y})\over |p_{3\perp}|}\, .
\eeqa

\section{Axial matrix elements in canonical representation}
\label{app:axialinst}
\subsection{Evaluation of $\bra{\ohalf\ohalf}\,A_+\, \ket{\ohalf,-\ohalf} $}
\beqa
&&\bra{\ohalf\ohalf}\, A_+\, \ket{\ohalf,-\ohalf} \,=\, 
-g_A^q \; {3\over2} \; \; {\cal N} \; {1\over(E_1'+m)(E_1+m)}
\Bigg\{
[ (E_1'+m)(E_1+m)-p_{1z}'p_{1z}] \nonumber \\
&&
\left[
\cos{\beta'\over2} \cos{\beta\over2} \; {\cal A} 
-
{2\over9} \sin{\beta'\over2} \sin{\beta\over2} {p_{1N+}p_{1N+} \over  |p_{\perp}|^2} \; {\cal B}
+ {1\over9}
\sin \left({\beta -\beta'\over2}\right) 
{p_{1N+}\over |p_{1\perp}|} \; {\cal C} \right] 
\nonumber\\
&&+
{\cal A}\; \left\{ |p_{1\perp}|^2 \; 
\left[ - \sin{\beta'\over2} \sin{\beta\over2}  \right]
-
{1\over2} [p_{1z}'+ p_{1z} ]
\sin{\left({\beta+\beta'\over 2}\right)} |p_{\perp}| \right\} \nonumber \\
&&- {2\over9} \;
{\cal B}\;p_{1N+}p_{1N+}\;  \left [
-\cos{\beta'\over2} \cos{\beta\over2}  
+{1\over2}
[p_{1z}'+ p_{1z} ]
\sin{\left({\beta+\beta'\over 2}\right)} {1\over |p_{\perp}|}\; \right] \nonumber \\
&&-
{1\over9}\; {\cal C}\; \left (
 \cos{\beta'\over2} \sin{\beta\over2} 
-
\sin{\beta'\over2}\cos{\beta\over2} \right ) 
p_{1N+}|p_{1\perp}|  \nonumber \\
&&+
{1\over2} (p_z'-p_z)
\left[ {2\over9}
\cos{\left({\beta'-\beta\over 2}\right)} \;p_{1N+}\;{\cal C}
\right. \nonumber \\
&&+ \left.
\sin{\left({\beta'-\beta\over 2}\right)} {1\over|p_{1\perp}|}
\left[
|p_{1\perp}|^2 \;
{\cal A}-{2\over9}p_{1N+} p_{1N+}\; {\cal B} \right]\right]
\Bigg\}\, ,
\eeqa
where, $p_{iN\pm}=p_{ix}\pm i p_{iy}$,
\beqa
{\cal N}=\sqrt{E_1'+m\over 2 E_1'}\sqrt{E_1+m\over 2 E_1} \, ,&\qquad&
{\cal A}= {2\over 9} [5\,c_2c_3+4\, \hat{p}_{2\perp}\cdot \hat{p}_{3\perp}] \, ,
\nonumber \\
{\cal B}=\;s_2s_3\;{p_{2N-}p_{3N-} \over|p_{2\perp}||p_{3\perp}|} \, ,&\qquad&
{\cal C}=[ c_2( a_3+ i b_3) + c_3(a_2+ib_2) ]\, , 
\eeqa
and, 
\beqa
&&\beta: =\theta_1  \,,\qquad c_i=\cos{{\theta_i'-\theta_i \over 2}} \, ,
\qquad  s_i=\sin{{\theta_i'-\theta_i \over 2}} \, , \nonumber\\
&&\beta':=\theta'_1 \,,\qquad a_i ={p_{ix}\over |p_{i\perp}|} s_i \, ,  
\qquad b_i=-{p_{iy}\over |p_{i\perp}|} s_i \, .
\eeqa
The angles $\theta_i$ have already been defined in Eq.~(\ref{THETA}).

\subsection{Evaluation of $\bra{\ohalf\ohalf}\, A_z \,\ket{\ohalf\ohalf}$ }
\beqa
&&\bra{\ohalf\ohalf} \,  A_z \,  \ket{\ohalf\ohalf}
=
-g_A^q 
\sqrt{E_1'+m\over 2 E_1'}
\sqrt{E_1+m\over 2 E_1}
\bra{p\uparrow} 3 \tau^1_z \nonumber \\
&&\left\{ \sigma^1_z \left[ \left( 1-{|\vec{p}_\perp|^2 -p_{1z}p_{1z}' \over (E_1'+m)(E_1+m)}\right)
-{m (p_1'-p_1)_z \over (\vec{p_1'}-\vec{p_1})^2 +m_\pi^2}
\left({ p_{1z}' \over E_1'+m}
-{ p_{1z}\over E_1+m} \right ) \right] \right. \nonumber \\
&&+
\sigma_{1+} p_{1N-} \left[ {p_{1z}+p_{1z}' \over  (E_1'+m)(E_1+m)} 
-{m (p_1'-p_1)_z \over (\vec{p_1'}-\vec{p_1})^2 +m_\pi^2}
{E_1'-E_1 \over (E_1'+m)(E_1+m)} \right] \nonumber \\
&&+ \left.
\sigma_{1-} p_{1N+}  \left[ {p_{1z}+p_{1z}' \over  (E_1'+m)(E_1+m)} 
-{m (p_1'-p_1)_z \over (\vec{p_1'}-\vec{p_1})^2 +m_\pi^2}
{E_1'-E_1 \over (E_1'+m)(E_1+m)} \right] \right\}  \ket{p\uparrow}\ .
\eeqa
One may define ${\cal D}$ so that it becomes,
\beqa
\bra{\ohalf\ohalf} A_z(Q^2)  \ket{\ohalf\ohalf}
&&= 
-g_A^q 
{\cal N}
\bra{p\uparrow} 3 \tau^1_z 
\left\{ \sigma^1_z \; {\cal D}_z
+ \sigma_{1+} p_{1N-}  \; {\cal D}_\perp 
+ \sigma_{1-} p_{1N+}  \; {\cal D}_\perp 
\right\}  \ket{p\uparrow}\, .
\eeqa

The final result can be expressed as,

\beqa
 \langle A \rangle_z &=& 
-g_A^q \; {3\over2} \; 
{\cal N} \; {\cal D}_\perp {1\over 9}
\left\{
\cos{\beta'\over2} \cos{\beta\over2} 
\left[ c_2s_3 {p_{3N+}p_{1N-}\over |p_{3\perp}|} 
      +c_3s_2 {p_{2N+}p_{1N-}\over |p_{2\perp}|} \right]  \right.\nonumber \\
&+& 
\sin{\beta'\over2} \sin{\beta\over2} 
\left[ c_2s_3{p_{3N-}p_{1N+}\over |p_{3\perp}|} 
      +c_3s_2{p_{2N-}p_{1N+}\over |p_{2\perp}|} \right]
\nonumber \\
&-&\left.
\cos{\beta'\over2} \sin{\beta\over2} 
|p_{1\perp}| \;
[c_2c_3 -10 s_2s_3 \hat{p}_{2\perp}\cdot \hat{p}_{3\perp}]
+
2 \,  \sin{\beta'\over2}\cos{\beta\over2} 
|p_{1\perp}|   \;c_2c_3  
\right\} \nonumber \\
&-&
g_A^q \; {3\over2}  \; 
{\cal N}{\cal D}_z  {1\over 9}\left\{
\cos{\left({\beta+\beta'\over 2}\right)} 
\left[c_2c_3 +10 s_2s_3 \hat{p}_{2\perp}\cdot \hat{p}_{3\perp}\right]\right. \nonumber \\
&+& 
\sin{\left({\beta+\beta'\over 2}\right)} 
{1\over |p_{\perp}|}
\left( c_2s_3{p_{3N-}p_{1N+}\over |p_{3\perp}|} 
                +c_3s_2{p_{2N-}p_{1N+}\over |p_{2\perp}|}  \right. \nonumber\\
&-&\left.\left.
                 c_2s_3 {p_{3N+} p_{1N-} \over |p_{3\perp}|} 
               -c_3s_2{p_{2N+} p_{1N-} \over |p_{2\perp}|}
\right ) \right\} \nonumber \\
&-&
g_A^q \; {3\over2} \; 
{\cal N} {\cal D}_\perp \,{1\over 9}
\left\{
-\cos{\beta'\over2} \cos{\beta\over2} 
\left[ c_2s_3{p_{3N-} p_{1N+}\over |p_{3\perp}|} 
      +c_3s_2{p_{2N-} p_{1N+}\over |p_{2\perp}|} \right] \right.\nonumber \\
&-&
\sin{\beta'\over2} \sin{\beta\over2} 
\left[ c_2s_3 {p_{3N+}p_{1N-}\over |p_{3\perp}|} 
      +c_3s_2{p_{2N+}p_{1N-}\over |p_{2\perp}|} \right] 
\nonumber \\
&+& \left.
2\,
\cos{\beta'\over2} \sin{\beta\over2} 
|p_{1\perp}|  \,c_2c_3   
-
\sin{\beta'\over2}\cos{\beta\over2} 
|p_{1\perp}| 
[c_2c_3 -10 s_2s_3 \hat{p}_{2\perp}\cdot \hat{p}_{3\perp}]
\right\}\, , \nonumber
\eeqa

%\newpage
\section{Axial matrix elements in light front representation}
\label{app:af}
\subsection{Matrix element $\bra{\half} A^+ \ket{\half}$}

\beqa
&&\bra{ \half} A^+ \ket{\half} =
{3\over2}
g_A^q 
\left\{
{1\over D_1D_2D_3} 
[a'_1 a_1+i (\vec{q}_\perp \times\vec{q'}_\perp)_z -\vec{q'}_\perp \cdot \vec{q}_\perp]
[  f_2 f_3 +  \vec{V}_{2}\cdot\vec{V}_{3} ]
\right.\\
&-&
{1\over3}
 \left(
{1\over D_1D_2D_3} {1\over3}  
[a'_1 a_1+i (\vec{q}_\perp \times\vec{q'}_\perp)_z - \vec{q'}_\perp \cdot \vec{q}_\perp]
[ f_2 f_3 - V_{2x}V_{3x} - V_{2y}V_{3y} + V_{2z}V_{3z} ] 
\right.\nonumber \\
&+&{1\over D_1D_2D_3} {2\over3}  
[-a'_1 a_1+i (\vec{q}_\perp \times\vec{q'}_\perp)_z + \vec{q'}_\perp \cdot \vec{q}_\perp]
[ f_2 f_3 + if_2 V_{3z}  + if_3 V_{2z}  - V_{2z}V_{3z} ]\nonumber \\
&-&{1\over D_2D_3} {1\over3}  
[a'_1 (q_x-iq_y) +a_1 (q'_x-iq'_y)]
[(f_2+iV_{2z})( iV_{3x} - V_{3y}) + (f_3+iV_{3z}) (i V_{2x}- V_{2y})]
\nonumber \\
&-&\left. \left.
{1\over D_1D_2D_3} {1\over3}  
[a'_1 (q_x+iq_y) + a_1 (q'_x+iq'_y)]
[(f_2+iV_{2z})( iV_{3x} + V_{3y}) + (f_3+iV_{3z}) (iV_{2x} + V_{2y}) ]
\right ) \right\} \nonumber 
\eeqa
with the definitions of Eq.~(\ref{frontnot}).

\subsection{Matrix element $\bra{-\half} (A_x+iA_y) \ket{\half}$}
An approximate expression (with $V_{2i}=V_{3i}=0$, 
for the sake of clarity) is:
\beqa
\bra{-\half} (A_x+iA_y) \ket{\half} &\simeq& 
{3\over2} g_A^q {1\over D_1D_2D_3} (f_2 f_3) {10\over 9}
\nonumber \\
&&\bigg\{
\bigg(1-{Q^2\over Q^2+m_\pi^2}\bigg) 
(a'_1 a_1+q'_{1y}q_{1y}-q'_{1x}q_{1x}
-i(q_{1x}q'_{1y} + q_{1y}q'_{1x}))
\nonumber \\
&+&
\bigg({p_{1x}+ip_{1y}+p'_{1x}+ip'_{1y}\over 2m}\bigg)
(a'_1 ( q_{1x}+iq_{1y})+a_1 (q'_{1x}+iq'_{1y}) )\nonumber \\
&-&
(a'_1 a_1-q'_{1y}q_{1y}+q'_{1x}q_{1x})  \nonumber \\
&-&
(i V_{1x}-V_{1y}) { Q\over 2 m} \bigg\}
\eeqa
with the definitions of Eq.~(\ref{frontnot}).

\end{document}